\documentclass[twocolumn, 10pt]{iopjournal}

\usepackage{geometry}
\usepackage[utf8]{inputenc}
\usepackage[T1]{fontenc}
\usepackage{amsmath,amssymb}
\usepackage{amsthm}

\usepackage{bm}
\usepackage{bbm}
\usepackage{graphicx}
\usepackage{tikz}
\usetikzlibrary{calc}
\usepackage{mathtools}
\usepackage{breqn}
\usepackage{caption}
\usepackage{subcaption}
\usepackage{cite}
\usepackage{stmaryrd}
\usepackage{comment}
\usepackage{multicol}
\excludecomment{comment}

\newcommand{\panelgraphic}[3][]{%
  \begin{tikzpicture}[baseline=(panelimage.south)]
    \node[inner sep=0] (panelimage) {\includegraphics[#1]{#2}};
    \node[anchor=north west,fill=white,fill opacity=0.8,text opacity=1,inner sep=1pt]
      at ([xshift=14pt,yshift=-11pt]panelimage.north west) {\textbf{(#3)}};
  \end{tikzpicture}%
}
\newcommand{\compactpanelgraphic}[3][]{%
  \begin{tikzpicture}[baseline=(panelimage.south)]
    \node[inner sep=0] (panelimage) {\includegraphics[#1]{#2}};
    \node[anchor=north west,fill=white,fill opacity=0.8,text opacity=1,inner sep=0.5pt]
      at ([xshift=24pt,yshift=-8pt]panelimage.north west)
      {\scriptsize\textbf{(#3)}};
  \end{tikzpicture}%
}

\begin{document}
\onecolumn
\articletype{Paper} %	 e.g. Paper, Letter, Topical Review...

\title{Complex nonlinear dynamics of area-preserving, active vesicles}
\author{Reiner Kree$^1$\orcid{0000-0001-9500-7667} and Annette Zippelius$^{1,*}$\orcid{0000-0000-0000-0000}}

\affil{$^1$Institut f\"ur Theoretische Physik, Georg-August Universit\"at G\"ottingen, Germany}

\email{rkree1@phys.uni-goettingen.de}

\keywords{active vesicles; geometric swimming; nonlinear shape dynamics; low-Reynolds-number propulsion; membrane inextensibility}

%\date{\today}

    \begin{abstract}

We investigate the nonlinear shape dynamics and autonomous propulsion of
actively driven quasi-spherical vesicles with locally inextensible
membranes at low Reynolds number. Starting from Stokes hydrodynamics,
linearized membrane elasticity, and harmonic active forcing, we derive a reduced
description in terms of spherical harmonic deformation modes.
The global area constraint enforced by local inextensibility is the sole source of dynamic nonlinearity.
It confines the dynamics to compact manifolds in
the space of possible shapes. Autonomous propulsion arises through nonlinear mode
coupling and is determined geometrically by the oriented area swept by
the  trajectories in shape space.
For two active modes, the dynamics reduces to a periodically driven
phase equation exhibiting synchronization, phase slips, and mode
locking. Introducing a third active mode fundamentally changes the
dynamics, giving rise to quasiperiodic invariant tori and resonant
periodic cycles. A recurrence diagnostic reveals the resulting resonance
structure, while fluctuations of the cycle-averaged propulsion provide
an experimentally accessible signature of the underlying shape dynamics.
Our results demonstrate that, for actively driven vesicles, a geometric constraint is sufficient to transform an otherwise linear dynamical system into one exhibiting rich nonlinear dynamics. 
\end{abstract}

\twocolumn
\section{Introduction}
At low Reynolds number, inertia is negligible, and locomotion therefore requires non-reciprocal shape changes, as expressed by Purcell’s scallop theorem~\cite{Purcell1977}. Taylor's pioneering analysis of an
inextensible waving sheet provided an early realization of this
mechanism and revealed the nontrivial kinematic consequences of local
inextensibility~\cite{Taylor1951}. Lighthill~\cite{Lighthill1952} and
Blake~\cite{Blake1971} subsequently analyzed propulsion by deformable
nearly spherical bodies. Shapere and Wilczek later formulated the
locomotion of deformable swimmers as geometric swimming: the net
displacement is determined by the path traced through the space of
possible shapes~\cite{ShapereWilczek1987,ShapereWilczek1989,
KellyMurray1995,StoneSamuel1996}.

Fluid membrane vesicles provide a particularly rich setting in which
to study hydrodynamic and geometric swimming and serve as paradigmatic models for red
blood cells, giant vesicles, protocells, and active biomimetic
systems~\cite{Helfrich1973,Seifert1997,FarutinMisbahPeyla2011EPJE,Deserno2015}.
Their membranes resist bending while enclosing a fixed volume and
conserving local membrane area. Local inextensibility couples normal
deformations to tangential membrane flows, while the resulting global
area constraint couples the deformation modes. The reduced shape
dynamics can therefore be nonlinear even when the hydrodynamic
response and membrane forces are linearized.

Activity may enter through several physically distinct mechanisms.
Spatial variations of spontaneous curvature or membrane elastic
properties~\cite{SensTurner2006}, as well as tractions exerted by
submembrane cortices or active gels~\cite{Gov2006,Alimohamadi2018}, can
drive autonomous or externally controlled shape changes. These mechanisms can now be studied experimentally
using reconstituted actomyosin
cortices~\cite{Carvalho2013,Loiseau2016}, membrane reaction--diffusion
systems~\cite{Loose2011PNAS,Litschel2018}, and optogenetic or light-driven
membrane remodeling~\cite{Pernpeintner2017_Langmuir,Jones2020}. In
particular, encapsulated Min-protein oscillations can drive autonomous,
reversible shape oscillations of giant
vesicles~\cite{Litschel2018,Christ2021}, demonstrating that internally generated
biochemical oscillations can act as time-dependent sources of membrane
deformation.

Previous theoretical studies have established that time-dependent
membrane forcing can produce self-propulsion.
~\cite{FarutinMisbahPeyla2011EPJE,Farutin2013,Kree2025}. For a
deformable swimmer, however, the forcing does not directly prescribe a
swimming stroke. Instead, it generates a trajectory in shape space
through the membrane mechanics and constraints, and the geometry of
this trajectory determines the resulting displacement. When the shape
response is nonlinear, relating active forcing to propulsion therefore
requires a detailed analysis of the shape dynamics. We use concepts and methods 
from nonlinear dynamics to establish this link for periodically driven active vesicles.

Our starting point is  the force balance between hydrodynamic stresses, passive
membrane forces, and active forces generated by membrane and cortical
processes. We show how spatially and temporally varying spontaneous
curvature, bending rigidity, and Gaussian rigidity, together with
membrane--cortex couplings, enter the theory through effective active
membrane tractions. The temporal forcing protocols are prescribed, but
the resulting deformation histories are determined dynamically rather
than imposed as paths in shape space, as in kinematic swimming
theories~\cite{ShapereWilczek1987,ShapereWilczek1989,KellyMurray1995}.

To obtain an analytically tractable description, we consider weakly
deformed vesicles in the quasi-spherical
approximation~\cite{MilnerSafran1987}. We derive a dynamical system for
the deformation modes in which the area constraint confines the
evolution to compact manifolds in shape space. Although the
hydrodynamic response, membrane mechanics, and external forcing are
individually linear, this constraint renders the reduced shape dynamics
nonlinear. Combining these equations with the kinematic propulsion law
in the form derived previously~\cite{Kree2025}, we determine how the
dynamically generated trajectory in shape space produces the swimming
velocity. Propulsion thereby becomes a geometric observable of the
nonlinear shape dynamics.

Our analysis demonstrates this connection in the minimal two- and
three-mode truncations. The two-mode system reduces to a periodically
driven phase equation. Propulsion sets in through a saddle-node
bifurcation on an invariant circle, and its mean value is controlled by
the rotation number. Retaining a third deformation mode permits
dynamically selected invariant tori and resonant cycles. Moreover,
unlike the two-mode system, the three-mode system already produces
finite mean propulsion in the weakly forced synchronized regime. In
this higher-dimensional setting, statistics of the cycle-resolved
propulsion reveal dynamical transitions that are only weakly visible in
the mean velocity.

\section{Geometry, membrane mechanics and hydrodynamics of active vesicles}
\label{sec:generalformulation}

\subsection{Geometry and kinematics}

We consider a vesicle bounded by a closed fluid membrane and immersed in
an incompressible Newtonian fluid at low Reynolds number. The vesicle
encloses an internal fluid of viscosity $\eta^{-}$ and is surrounded by
an external fluid of viscosity $\eta^{+}$. Neglecting inertia, the
velocity fields $\bm v^{\pm}$ in the inner ($-$) and outer ($+$) fluids
satisfy the Stokes equations
\cite{KimKarrila2005,Seifert1997}, 
\[
\nabla\!\cdot\bm{\sigma}^{\pm}=0,
\qquad
\nabla\!\cdot\bm v^{\pm}=0,
\]
with Newtonian stress tensors
\[
\sigma^\pm_{ij}
=
\eta^\pm
\left(
\partial_i v^\pm_j+\partial_j v^\pm_i
\right)
-p^\pm\delta_{ij}.
\]
The laboratory frame is chosen as the rest frame of the undisturbed
external fluid.

We restrict attention to star-shaped vesicles, whose membrane is represented as a radial graph about a time-dependent reference point $\bm R(t)$~\cite{Seifert1997, MilnerSafran1987}.
In the laboratory frame, the membrane position is thus written as $\bm X(\theta,\phi,t)=\bm R(t) + \bm r_s(\theta,\phi,t)$, with
\begin{equation}
\bm r_s(\theta,\phi,t)=
a\bigl[1+f(\theta,\phi,t)\bigr]\bm e_r(\theta,\phi),
\label{eq:shape_map}
\end{equation}
where  $(\theta,\phi)$ denote Eulerian coordinates on a reference sphere of radius $a$ centered at  $\bm R(t)$.

The radius $a$ is  defined by the conserved vesicle volume,
$
V=4\pi a^3/3,
$
so that the reference sphere has the same volume as the vesicle.
For simplicity, we restrict the present analysis to axisymmetric shapes
and forcing without swirl, so that $f$ is independent of $\phi$.
We therefore expand
\begin{equation}
f(\theta,t)=\sum_{\ell}f_\ell(t)\,P_\ell(\cos\theta),
\label{eq:deformation_Legendre_expansion}
\end{equation}
where $P_\ell$ are Legendre polynomials.

The decomposition of membrane motion into a rigid translation and a shape deformation is not uniquely determined by equation~\eqref{eq:shape_map}. The ambiguity arises from the choice of reference point used to characterize the rigid-body motion, which we will refer to as the translational gauge. Throughout this work we use the \emph{center-of-deformation} (CoD) frame, whose origin is the time-dependent position $\bm R(t)$ in equation~\eqref{eq:shape_map} and  it is fixed by the gauge condition
\[
f_1(t)=0.
\]
In other words, the dipolar ($\ell=1$) mode is represented entirely by the rigid translation $\bm R(t)$, rather than by the deformation field.

\begin{figure}
\centering
\includegraphics[width=\columnwidth]{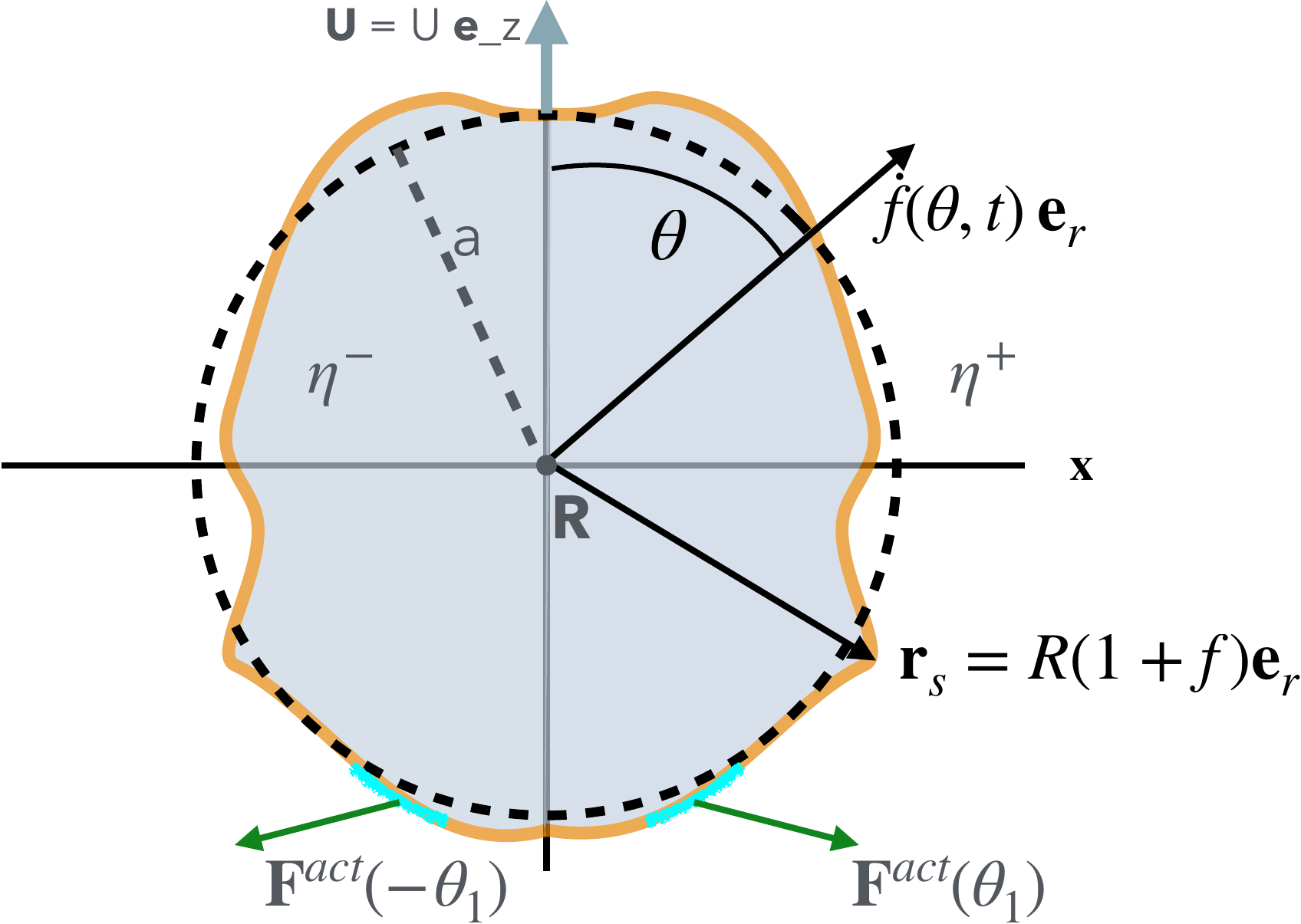}
\caption{
Schematic of an actively driven quasi-spherical vesicle of fixed volume $V=4\pi a^3/3$, separating inner and outer Newtonian fluids of viscosities
$\eta^{-}$ and $\eta^{+}$.
The membrane is subject to prescribed axisymmetric active traction
$\bm F^{act}(\theta,t)$ containing normal and tangential components while
preserving local inextensibility.
The instantaneous shape is represented as a radial graph
$\bm r_s(\theta,t)=a[1+f(\theta,t)]\bm e_r$ about the reference sphere
$r=a$.
The resulting force- and torque-free motion generates a swimming
velocity $\bm U=U\bm e_z$.
The dashed curve indicates the undeformed reference sphere.
}
\label{fig:intro_figure}
\end{figure}

The material surface velocity may be decomposed into normal and tangential components,
\[
\bm V=V_n\bm n+\bm V_t,
\]
where $\bm n$ denotes the outward unit normal and $\bm V_t$ the tangential membrane flow.
We employ an Eulerian description, which follows the evolution of the membrane as a geometric surface rather than tracking individual material points.
Consequently,
\[
\bm V
=
\frac{d\bm X(\theta(t),t)}{dt}
=
\partial_t\bm X+\dot\theta\,\partial_\theta\bm X,
\]
where an overdot denotes the time derivative. The second term  is tangential and represents the motion of the surface coordinates. It should not be identified with the material tangential velocity $\bm V_t$. Projecting onto the normal direction yields the kinematic relation in the CoD frame,
\begin{equation}
a(\bm e_r\!\cdot\bm n)\,\partial_t f=V_n.
\label{eq:kinematic}
\end{equation}
As anticipated, tangential flows along the membrane do not directly modify its shape. 
The membrane acts as a no-slip boundary, so that the bulk fluid velocity
coincides with the material surface velocity,
\[
\bm v^\pm(\bm X)=\bm V.
\]
In the CoD frame, the translational velocity
\(
\bm U=\dot{\bm R}
\)
appears as a uniform flow at infinity. Accordingly, the outer flow
satisfies
\begin{equation}
    \bm v^+(\bm r)\to-\bm U,
\qquad r\to\infty.
\label{eq:infinity_bc}
\end{equation}

\subsection{Inextensibility and excess area}

The membrane is locally inextensible. Consequently, the
surface velocity satisfies the kinematic constraint~\cite{Seifert1997, ArroyoDeSimone2009}
\begin{equation}
\nabla_s\cdot \bm V=\nabla_s\!\cdot\bm V_t+H\,V_n=0,
\label{eq:inextensibility}
\end{equation}
where $\nabla_s\!\cdot$ denotes the surface divergence and $H$ is the mean
curvature, defined such that $H=2/a$ for the reference sphere. For incompressible fluids this relation is equivalent to $\partial_n v^\pm_n=0$ on the membrane surface, as shown in
appendix~\ref{app:lamb_tractions}. 

For axisymmetric flows without swirl, the tangential velocity can be
written as a surface gradient of a scalar potential,
\begin{equation}
\bm V_t=\nabla_s V_t,
\label{eq:grad_V_t}
\end{equation}
so that equation~\eqref{eq:inextensibility} reduces to a surface Poisson
equation for $V_t$. Together with the kinematic relation
\eqref{eq:kinematic}, this determines the tangential membrane flow
uniquely from the shape velocity $\dot f$. Hence, $V_t$ is not an
independent degree of freedom.
Since both the kinematic relation \eqref{eq:kinematic} and the
inextensibility constraint \eqref{eq:inextensibility} are linear in the
velocities, the tangential membrane flow depends linearly on the shape
velocity $\dot f$. 

Local inextensibility implies conservation of the total membrane
area,
\begin{equation}
A_0=4\pi a^2(1+\Delta),
\end{equation}
where $\Delta$ denotes the conserved excess area. It limits the
admissible deformation amplitudes. 

\subsection{Membrane forces and active driving}

The membrane deformation is determined by local force balance, which
relates the jump of hydrodynamic tractions to passive membrane tractions
and active tractions,
\begin{equation}
\llbracket\bm\sigma\cdot\bm n\rrbracket
=
\bm F^{\rm mem}
+
\bm F^{\rm act},
\label{eq:traction_balance}
\end{equation}
where
\[
\llbracket\bm\sigma\cdot\bm n\rrbracket
=
(\bm\sigma^+-\bm\sigma^-)\cdot\bm n.
\]

Because the Stokes equations are linear, the hydrodynamic tractions
exerted by the inner and outer fluids are linear functionals of the
surface velocity~\cite{KimKarrila2005}. Through the kinematic relation
and local inextensibility, the surface velocity is itself determined by
the shape velocity $\dot f$. In the CoD frame we may therefore write
schematically
\begin{equation}
\bm F^\pm_{\rm hydro}
=
\hat{\bm T}^\pm[f]\dot f,
\end{equation}
where the linear operators $\hat{\bm T}^\pm[f]$ depend parametrically on
the instantaneous shape. Their explicit form for a spherical reference
state is derived in appendix~\ref{app:lamb_tractions}.

Substituting the hydrodynamic response into
equation~\eqref{eq:traction_balance} gives the symbolic evolution equation
\begin{equation}
\big(\hat{\bm T}^+[f]
-
\hat{\bm T}^-[f]\big)\dot f
=
\bm F^{\rm mem}[f]
+
\bm F^{\rm act},
\label{eq:shape_symbolic}
\end{equation}
which determines the membrane deformation once the passive and active
tractions are specified.

We distinguish two sources of tractions. The first originates
from membrane processes, while the second consists of direct
tractions exerted by an active sub-membrane cortex.
The passive membrane tractions are described within the Helfrich
model~\cite{Helfrich1973,Deserno2015}, with free energy
\begin{equation}
\mathcal F=
\int_S
\left[
\frac{\kappa}{2}(H-C)^2+\kappa_GK+\gamma
\right]\,dA,
\label{eq:Helfrich}
\end{equation}
where $H$ and $K$ denote the mean and Gaussian curvatures,
$\kappa$, $\kappa_G$, and $C$ may in general vary over the membrane,
representing the bending modulus, Gaussian modulus, and spontaneous
curvature, respectively, while $\gamma$ is the surface tension field
enforcing local membrane inextensibility (see appendix~\ref{app:helfrich_linear} for further details).
Uniform material parameters give the passive
membrane tractions $\bm F^{mem}$, whereas prescribed time-dependent spatial variations of the
spontaneous curvature and elastic moduli generate the active membrane
tractions, providing an effective description of the active membrane processes~\cite{Pernpeintner2017_Langmuir,Jones2020,Loose2011PNAS,Litschel2018} discussed in the Introduction.
 The cortical contribution enters separately as part of the active traction $\bm F^{\rm act}$ in
equation~\eqref{eq:traction_balance}.
Since both membrane and cortical activity arise from internal processes,
they generate no net force,
\begin{equation}
\int_S
\bm F^{\rm act}\,dA
=
\bm0,
\label{eq:force_free}
\end{equation}
and, for the axisymmetric no-swirl setting considered here, no net
torque.

\subsection{Propulsion velocity}

The local force balance determines the membrane deformation through the
shape dynamics derived above. The remaining unknown is the rigid
translation velocity $\bm U$, which is fixed by the global force-free
condition for the vesicle. Since membrane and cortical forces are
internal, the total external force is purely hydrodynamic. Integrating
the hydrodynamic traction over the membrane therefore yields
\begin{equation}
\bm0
=
\int_{S(t)}
\bm F^\pm_{\rm hydro}[\bm U+\bm V]\,dA.
\end{equation}

Decomposing the membrane velocity into a rigid translation and a
deformation velocity, and using the linearity of the Stokes equations,
the total hydrodynamic force separates into a contribution from rigid
translation and one due to the shape dynamics~\cite{StoneSamuel1996},
\begin{equation}
\bm0
=
\hat{\bm\zeta}^\pm\bm U
+
\int_{S(t)}
\hat{\bm T}^\pm[f]\dot f\,dA,
\label{eq:U_general_compact}
\end{equation}
where $\hat{\bm\zeta}^\pm$ denotes the hydrodynamic resistance tensor obtained by integrating the hydrodynamic traction associated with a rigid-body translation.

For axisymmetric motion,
$
\bm U=U\bm e_z,
$
so the resistance tensor reduces to a scalar drag coefficient.
Equation~\eqref{eq:U_general_compact} therefore determines the swimming
velocity uniquely once the shape dynamics has been obtained.

\section{Small--deformation expansion}
For quasi-spherical shapes, the coupled free-boundary problem described in the previous sections admits a controlled expansion that reduces it to a hierarchy of linear Stokes problems posed on a fixed reference sphere. The expansion parameter is determined by the excess area
\[
\Delta \ll 1,
\]
which controls the magnitude of the deformation amplitudes.

\subsection{Excess area and geometric constraints}

Expanding the membrane area and enclosed volume in powers
of the deformation field \(f\) yields
\begin{align}
\frac{A}{4\pi a^2}
&=
1+\Delta
\nonumber\\
&=
1
+
2\langle f\rangle
+
\left\langle
f^2+\frac12|\nabla f|^2
\right\rangle
+
O(f^3),
\label{eq:A_expand}
\\[1ex]
\frac{V}{4\pi a^3/3}
&=
1
+
3\langle f\rangle
+
3\langle f^2\rangle
+
O(f^3).
\label{eq:V_expand}
\end{align}
where 
\[
\langle \cdot \rangle
=
\frac1{4\pi}\int (\cdot)\,{\rm d}\Omega
\]
denotes the angular average over the unit sphere.
The fixed-volume constraint determines the mean
deformation \(\langle f\rangle\), corresponding to the
\(\ell=0\) mode, up to quadratic order.
Eliminating $\langle f\rangle$ from equation~\eqref{eq:A_expand} then yields the leading-order global area constraint
\begin{equation}
\Delta
=
\sum_{\ell\ge2}
w_\ell f_\ell^2
+
O(f^3),
\quad
w_\ell
=
\frac{(\ell-1)(\ell+2)}
{4\ell+2}.
\label{eq:area_constraint_modes}
\end{equation}

Equation~\eqref{eq:area_constraint_modes} implies the scaling
\[
f_\ell=O(\epsilon),
\qquad
\epsilon=\sqrt{\Delta},
\qquad
\ell\ge2,
\]
whereas the volume constraint requires the isotropic mode to satisfy
$
f_0=O(\epsilon^2).
$

\subsection{Rescaled mode variables and constrained dynamics}

It is convenient to rescale the deformation
amplitudes according to
\begin{equation}
q_\ell
=
\sqrt{\frac{w_\ell}{\Delta}}\,f_\ell,
\label{eq:q_def}
\end{equation}
in terms of which the excess-area constraint takes the form
\begin{equation}
\sum_{\ell\ge2} q_\ell^2
=
1
+
O(\epsilon).
\label{eq:unit_sphere_constraint}
\end{equation}

To leading order, the admissible deformations are therefore
confined to the surface of a unit sphere in the rescaled
mode space.

\subsection{Perturbation hierarchy on the reference sphere}
In the small-deformation expansion, all geometric quantities, boundary conditions, and force balances are systematically mapped from the deforming membrane to the fixed reference sphere.
To this end, every field is expanded in powers of the deformation amplitude,
\[
(\cdot)
=
(\cdot)^{(0)}
+
\epsilon(\cdot)^{(1)}
+
\epsilon^2(\cdot)^{(2)}
+
O(\epsilon^3).
\]
Boundary conditions originally posed on the deformed
surface
\[
r=a(1+f)
\]
are pulled back to the reference sphere \(r=a\) by a Taylor
expansion in the normal displacement.
In particular, the bulk velocity fields satisfy
\begin{align}
\bm V
&=
\bm v^\pm(r(f),\theta)
=
\bm v^\pm(a,\theta)
\nonumber\\
&
+
a f(t,\theta)
(\partial_r\bm v^\pm)(a,\theta)
+\cdots .
\label{eq:taylor_pullback}
\end{align}

Substituting the perturbation expansions of the bulk velocity fields
into equation~\eqref{eq:taylor_pullback} and collecting terms of equal order
in $\epsilon$ shows that each perturbative order receives contributions
both from the bulk fields at the same order and from the pullback of
lower-order solutions. Consequently, every order is governed by a linear
Stokes problem posed on the reference sphere.

% \section{Constrained shape dynamics}

% We now derive the leading-order evolution equations for the deformation amplitudes. Applying the small-deformation expansion to the kinematic relation~\eqref{eq:kinematic}, the local inextensibility~\eqref{eq:inextensibility}, and the force balance \eqref{eq:traction_balance} yields a closed set of evolution equations for the rescaled deformation mode amplitudes. 

\section{Constrained shape dynamics}

We now derive the leading-order evolution equations for the deformation
amplitudes. Applying the small-deformation expansion to the kinematic
relation~\eqref{eq:kinematic}, local
inextensibility~\eqref{eq:inextensibility}, and the force
balance~\eqref{eq:traction_balance} yields a closed set of evolution
equations for the rescaled deformation-mode amplitudes.

From this point onward, we use Legendre expansions not only for the
deformation but for all axisymmetric fields. Scalar fields are expanded
like $f(\theta)$ in equation~\eqref{eq:deformation_Legendre_expansion}, whereas 
axisymmetric vector fields like a velocity $\bm v$ are expanded in the form
$\bm v=v_r\bm e_r+v_\theta\bm e_\theta$ with
\begin{equation}
\begin{aligned}
v_r(\theta,t)
& =
\sum_{\ell\geq 0}v_{r,\ell}(t)P_\ell(\cos\theta),\\
v_\theta(\theta,t)
&=
\sum_{\ell\geq 1}v_{\theta,\ell}(t)
\partial_\theta P_\ell(\cos\theta).
\end{aligned}
\end{equation}

\subsection{Linearized hydrodynamic response}

At first order, the normal and tangential surface
velocities reduce to
\[
V_n^{(1)}=V_r^{(1)},
\qquad
\bm V_t^{(1)}=\nabla_0 V_\theta^{(1)},
\]
where $\nabla_0$ denotes the surface gradient on the unit
sphere and the second equation follows from  equation~\eqref{eq:grad_V_t}.

To leading order, the hydrodynamic traction-jump operators introduced in
Section~\ref{sec:generalformulation} are evaluated on the spherical reference state.
In the Legendre basis the projections in normal and tangential directions are diagonal, giving
\begin{align}
\bm e_r\!\cdot\!\llbracket\bm\sigma\!\cdot\!\bm n\rrbracket
&=
-\sum_{\ell\ge1}
\eta^+N_\ell\dot f_\ell P_\ell,
\label{eq:traction_normal}
\\
\bm e_\theta\!\cdot\!\llbracket\bm\sigma\!\cdot\!\bm n\rrbracket
&=
-\sum_{\ell\ge1}
\eta^+T_\ell\dot f_\ell
\partial_\theta P_\ell.
\label{eq:traction_tangential}
\end{align}
The corresponding eigenvalues are 
\[
N_\ell=
\frac{\lambda}{\ell}(\ell-1)(2\ell+3)
+
\frac{1}{\ell+1}(\ell+2)(2\ell-1),
\]
and
\[
T_\ell=
\frac{\lambda(\ell-1)+(\ell+2)}
{\ell(\ell+1)},
\qquad
\lambda=\eta^-/\eta^+.
\]
The coefficients $N_\ell$ and $T_\ell$ characterize the linear
hydrodynamic response of a spherical vesicle to normal and tangential
surface motions, respectively.
Their derivation is given in appendix~\ref{app:lamb_tractions}. 

\subsection{Linearized membrane dynamics}

We next linearize the membrane tractions entering the force
balance~\eqref{eq:traction_balance}. Details of the derivation are given in
appendix~\ref{app:helfrich_linear}.
The material parameters are decomposed into homogeneous reference
values and small, prescribed inhomogeneities,
$
M(\bm X,t)=M_0+\delta M(\bm X,t),
$
with
$
M\in\{\kappa,C,\kappa_G\}.
$
The homogeneous values determine the passive reference state and its
elastic response, whereas the inhomogeneous contributions enter the
linearized mode equations as active forcing. The membrane tension is
decomposed analogously
$
\gamma(\bm X,t)=\gamma_0+\delta\gamma(\bm X,t).
$
Unlike the material parameters, however, the tension is not prescribed
but acts as a Lagrange-multiplier field. Its nonuniform part
$\delta\gamma$ enforces local inextensibility, while its uniform part
$\gamma_0$ is determined by conservation of the total membrane area.
We use the vesicle radius $a$ as the characteristic length and
\[
\tau_b=\frac{\eta^+a^3}{\kappa_0}
\]
as the characteristic bending-relaxation time. From this point onward,
all quantities are dimensionless. In particular, the dimensionless
spontaneous-curvature mismatch is defined as
\[
m=2-aC_0.
\]

For $\ell\ge2$, the normal and tangential force balances
become
\begin{align}
- N_\ell \dot f_\ell
&=
g_\ell(\beta_\ell+\gamma_0)f_\ell
+
2\delta\gamma_\ell
+
F^{act}_{r,\ell},
\label{eq:balance_normal}
\\
- T_\ell \dot f_\ell
&=
-\delta\gamma_\ell
+
F^{act}_{\theta,\ell},
\label{eq:balance_tangential}
\end{align}
where
\[
g_\ell=(\ell+2)(\ell-1),
\qquad
\beta_\ell=g_\ell+m^2/2.
\]

The geometric factor $g_\ell$
 arises from the first variation of the surface area element and therefore appears universally in membrane forces derived from surface energy densities. The coefficient $\beta_\ell$ 
 represents the mode-dependent passive bending stiffness, including the effect of the spontaneous-curvature mismatch.

\subsection{Active forcing structure}

Eliminating the tension perturbation $\delta\gamma_\ell$
from equations~\eqref{eq:balance_normal} and~\eqref{eq:balance_tangential} yields the mode equation

\begin{equation}
\left(N_\ell+2T_\ell\right)\dot f_\ell
=
-g_\ell(\beta_\ell+\gamma_0)f_\ell
-
F^{act}_\ell ,
\label{eq:f_mode_dynamics}
\end{equation}
with
\[
F^{act}_\ell
=
F^{act}_{r,\ell}
+
2F^{act}_{\theta,\ell}.
\]
Equation~\eqref{eq:f_mode_dynamics} is the linearized mode representation of the evolution equation~\eqref{eq:shape_symbolic}. The continuum operators reduce to the scalar coefficients $N_\ell$, $T_\ell$
, while $F^{act}_\ell$ denotes the corresponding modal active forcing,
\begin{align}
F^{act}_\ell
&=
g_\ell
\left(
m\,\delta\kappa_\ell
-
\delta C_\ell
+
\delta\kappa_{G,\ell}
\right)
\nonumber\\
&\quad
+
\left(
F^{cortex}_{r,\ell}
+
2F^{cortex}_{\theta,\ell}
\right).
\label{eq:Fshape_explicit}
\end{align}
Its membrane-bound part originates from spatial variations of bending
rigidity, spontaneous curvature, and Gaussian rigidity, while the
remaining terms describe active cortical tractions (see appendix~\ref{app:helfrich_linear}).
An important structural property is that every membrane-bound
contribution carries the geometric factor $g_\ell$, which
vanishes for $\ell=1$. The global force-free condition
likewise removes the $\ell=1$ component of the cortical
forcing. As a result, the dipolar mode does not appear in
the shape dynamics.

Introducing the rescaled variables \eqref{eq:q_def} and multiplying equation~\eqref{eq:f_mode_dynamics} by $\sqrt{w_\ell/\Delta}$, we obtain
\begin{equation}
\left(N_\ell+2T_\ell\right)\dot q_\ell
=
-g_\ell(\beta_\ell+\gamma_0)q_\ell
-
\sqrt{\frac{w_\ell}{\Delta}}\,F^{act}_\ell .
\label{eq:q_component_dynamics}
\end{equation}
With rescaled forcing amplitudes, defined as
\[
g_\ell\,F_\ell=
\sqrt{w_\ell/\Delta}\,
F^{act}_\ell,
\]
the mode dynamics assumes the compact vector form
\begin{equation}
\dot{\bm q}
=
-
\hat{\bm M}^{-1}\hat{\bm g}
\left(
\hat{\bm\beta}\bm q
+
\gamma_0\bm q
+
\bm F
\right),
\label{eq:shape-dynamics}
\end{equation}
where $\bm F$ has components $F_\ell$, 
and the matrices
$\hat{\bm M}$, $\hat{\bm g}$, and $\hat{\bm\beta}$ are
diagonal with entries
$
M_\ell=N_\ell+2T_\ell,
$
$
g_\ell,
\beta_\ell,
$
respectively.

\subsection{Constraint geometry and projected flow}

In terms of the rescaled variables, the global area constraint \eqref{eq:unit_sphere_constraint} becomes
\[
|\bm q|^2=1,
\]
so that retaining $n$ modes confines the dynamics to the unit sphere in rescaled mode space.

It is convenient to introduce the effective
mobility tensor
\[
\hat{\bm\Gamma}
=
\hat{\bm M}^{-1}\hat{\bm g},
\]
which combines hydrodynamic mobility and geometric
coupling into a single anisotropic operator.

\begin{figure}
\includegraphics[width=0.75\columnwidth]{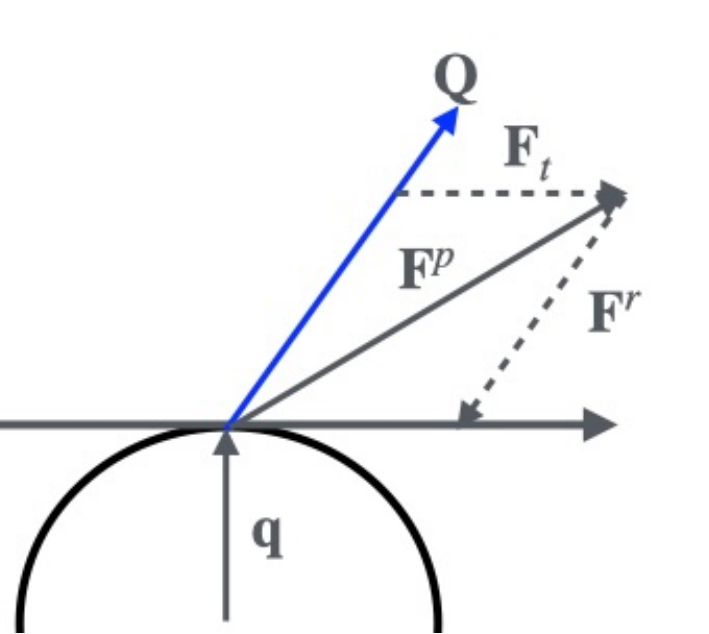}
\caption{Geometry of oblique projection.
The vector $\bm{F}^p$ is decomposed as $\bm{F}^p=-F^r\bm{Q} + F_{t} \bm q_t$, with reaction force $F^r$ and $\bm q\cdot \bm q_t=0$. The reaction force compensates the radial component of $\bm F^p$ and leaves the component $F_t\bm q_t = \hat{\bm P}\bm{F}^p $ in tangential space. }
\label{fig:oblique_projector}
\end{figure}

Defining
\[
\bm Q=\hat{\bm\Gamma}\bm q,
\qquad
\bm F^p
=
\hat{\bm\Gamma}\hat{\bm\beta}\bm q
+
\hat{\bm\Gamma}\bm F,
\]
Equation~\eqref{eq:shape-dynamics} can be written as
\[
\dot{\bm q}
=
-(\bm F^p+\gamma_0\bm Q).
\]
The homogeneous tension $\gamma_0$ 
 is determined by enforcing preservation of the global area constraint, equivalently requiring the dynamics to remain tangent to the constraint sphere,
$
\bm q\cdot\dot{\bm q}=0.
$
This yields
\begin{equation}
\gamma_0
=
-
\frac{\bm q\cdot\bm F^p}
{\bm q\cdot\bm Q}.
\label{eq:gamma_factorized}
\end{equation}
Substituting back gives the projected dynamics
\begin{equation}
\dot{\bm q}
=
-\hat{\bm P}\bm F^p,
\qquad
\hat{\bm P}
=
\left(
\hat{\bm I}
-
\bm Q
\frac{\bm q\cdot}
{\bm q\cdot\bm Q}
\right),
\label{eq:projectedDynamics}
\end{equation}
where $\hat{\bm P}$ is an oblique projector onto the
tangent space of the constraint sphere, as is shown in figure~\ref{fig:oblique_projector}. The projector removes the normal component of the generalized force incompatible with the area constraint, leaving only the admissible tangential motion.
Because the mobility tensor $\hat{\bm\Gamma}$ is generally
anisotropic, the projection is oblique rather than orthogonal.
The resulting evolution is therefore a constrained
relaxational flow on the unit sphere, driven by passive
elastic forces and active forcing.

\subsection{Mode truncation and reduced dynamics}

Although equation~\eqref{eq:shape-dynamics} is formally infinite-dimensional, the leading-order dynamics admits consistent finite-dimensional truncations.
For large $\ell$,
$
M_\ell\sim \ell,
$
and
$
g_\ell\beta_\ell\sim \ell^4.
$
Consequently, high-order deformation modes relax rapidly,  on the timescale
\[
\tau_\ell\sim \ell^{-3},
\]
and are only weakly excited by smooth forcing.
Furthermore, within the leading-order dynamics, any mode that is neither externally driven nor initially excited remains identically zero.
This justifies systematic low-dimensional reductions, in
particular the two-mode and three-mode truncations studied
below.

\subsection{Periodic forcing and time-glide symmetry}
\label{subsec:time_glide_symmetry}
Before analyzing particular mode truncations, we identify a discrete
symmetry of the projected dynamics for a broad class of periodic
drivings. These satisfy
\[
\bm F(t+T)=\bm F(t),
\qquad
\bm F(t+\vartheta\, T)=-\bm F(t),
\]
that is, the forcing reverses sign after a fixed fraction $\vartheta$ of
the forcing period. Harmonic forcing corresponds to the special case
$\vartheta=\tfrac12$, but the symmetry also encompasses more general
contraction--relaxation protocols and antagonistic activation patterns.

 . 

Because equation~\eqref{eq:projectedDynamics} is invariant under
simultaneous inversion of shape and forcing
the system  is  equivariant under the time-glide
transformation
\begin{equation}
(\bm q,t)
\mapsto
(-\bm q,t+\vartheta\, T),
\label{eq:symmetry}
\end{equation}

Thus, if \(\bm q(t)\) is a solution, then
$
-\bm q(t+\vartheta\, T)
$
is also a solution. 
Indeed, under the transformation
$(\bm q,t)\mapsto(-\bm q,t+\vartheta\, T)$,
both $\bm q$ and $\bm Q=\hat{\bm\Gamma}\bm q$ change sign, so the
projector $\hat{\bm P}$ remains unchanged, while the generalized
force $\bm F^p$ changes sign together with the forcing.
Invariant sets therefore occur either in
symmetry-related pairs or as self-symmetric states.

The time-glide symmetry admits a natural geometric interpretation in
the extended phase space of the periodically driven system, spanned by
the deformation modes $\bm q$ and the forcing phase
$
\varphi=\omega t,
\quad
\omega=2\pi/T,
$
which evolves according to
$
\dot\varphi=\omega.
$
In this space the symmetry transformation becomes
\[
(\bm q,\varphi)
\mapsto
(-\bm q,\varphi+2\pi\vartheta).
\]

\section{Propulsion from shape dynamics}

Having established the constrained mode dynamics, we now determine how the resulting trajectories in shape space generate translation of the vesicle. 
As shown in Ref.~\cite{Kree2026}, locally inextensible vesicles do not
propel at first order in the deformation amplitude. 
A non-vanishing translational velocity would require an
$\ell=1$ contribution to the surface velocity, originating from the
far-field boundary condition~\eqref{eq:infinity_bc}.
However, the linearized kinematic and local inextensibility
conditions admit no such mode in the CoD frame.
Consequently, self-propulsion first
appears at second order through nonlinear mode coupling~\cite{Lighthill1952}. 
Because the argument relies only on the linearized kinematic and local inextensibility conditions, it is independent of the microscopic origin of the active forcing.

In the remainder of this section, we summarize the second-order propulsion law derived in Ref.~\cite{Kree2025} and discuss its geometric interpretation together with its asymptotic limits for slow and fast driving.

\subsection{Second-order propulsion law}
The leading contribution to propulsion velocity arises at
$O(\varepsilon^2)$.
At this order, quadratic couplings between the deformation modes generate  net translation.
A key simplification is that the propulsion velocity can be expressed
entirely in terms of first-order quantities;
no second-order shape correction needs to be computed explicitly.

The second-order propulsion law was derived previously in
Ref.~\cite{ Lighthill1952, Blake1971, Kree2025}. In the present formulation, its structure
acquires a natural geometric interpretation.

For axisymmetric deformations, the propulsion velocity is directed along the symmetry axis, $\bm U = U \bm e_z$, and is given by
\begin{equation}
U
=
\sum_{\ell\ge 2}
\Big[
\mathcal{C}_\ell
\big(
q_\ell \dot q_{\ell+1}
-
q_{\ell+1}\dot q_\ell
\big)
+
B_\ell \frac{d}{dt}(q_\ell q_{\ell+1})
\Big],
\label{eq:U_decomposition}
\end{equation}
with
\begin{align}
\mathcal{C}_\ell  &=\frac{\Delta}{2D_\ell}(2\ell^2 + 6\ell -5),\\\nonumber
B_\ell &= \frac{\Delta}{2D_\ell}(2\ell^3 + 6\ell^2 +14\ell +11),
\end{align}
and
$D_\ell=(2\ell+1)(2\ell+3)\sqrt{w_\ell w_{\ell+1}}$.

Equation~\eqref{eq:U_decomposition} naturally separates into an antisymmetric and a symmetric contribution.
The symmetric contribution depends on the translational gauge, that is,
on the choice of reference point used to decompose the membrane motion
into rigid translation and deformation.
A shift of this point,
$
\bm R \rightarrow \bm R+\bm\chi(\bm q),
$
changes the instantaneous velocity according to
$
\bm U \rightarrow \bm U + d\bm\chi(\bm q)/dt.
$
Hence the symmetric term in equation~\eqref{eq:U_decomposition} is gauge dependent.
By contrast, the antisymmetric term is gauge invariant and has a direct geometric interpretation.
Indeed, adding a total time derivative modifies only the symmetric part of equation~\eqref{eq:U_decomposition}, leaving the antisymmetric bilinear form unchanged.

The gauge-invariant contribution to the propulsion velocity is therefore
determined by the antisymmetric combinations
\[
q_\ell\dot q_{\ell+1}
-
q_{\ell+1}\dot q_\ell.
\]
To each neighbouring pair of deformation modes we associate the signed
swept area
\begin{equation}
A_{\ell,\ell+1}(t)
=
\frac12
\int_0^t
\left(
q_\ell\,dq_{\ell+1}
-
q_{\ell+1}\,dq_\ell
\right),
\label{eq:area_definition}
\end{equation}
whose time derivative is
\[
\dot A_{\ell,\ell+1}
=
\frac12
\left(
q_\ell\dot q_{\ell+1}
-
q_{\ell+1}\dot q_\ell
\right).
\]
The quantity $A_{\ell,\ell+1}(t)$ is the signed area swept by the radius
vector joining the origin to the projected trajectory in the
$(q_\ell,q_{\ell+1})$ plane. Thus, propulsion depends on the local areal
motion of neighbouring deformation modes rather than on the individual
mode amplitudes.

Integrating equation~\eqref{eq:U_decomposition} over time yields the net
displacement
\begin{equation}
\begin{aligned}
\Delta Z(t)
&=
Z(t)-Z(0)
=
\sum_{\ell\ge2}
\Delta Z_\ell(t)
\\
&=
\sum_{\ell\ge2}
\left(
2\mathcal C_\ell A_{\ell,\ell+1}(t)
+
\Phi_\ell(t)
-
\Phi_\ell(0)
\right),
\end{aligned}
\label{eq:ZPropulsion}
\end{equation}
with
\[
\Phi_\ell(t)
=
B_\ell q_\ell(t)q_{\ell+1}(t).
\]

\begin{figure}
    \centering
    \includegraphics[width=\columnwidth]{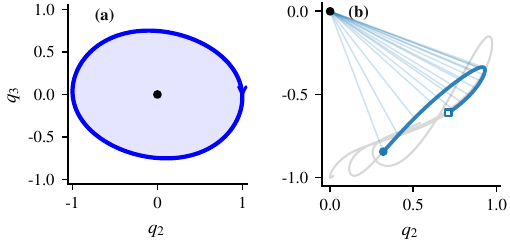}
\caption{Geometric interpretation of propulsion in adjacent mode planes. (a) Projection of a periodic trajectory in the $(q_2, q_3$
 plane. The net displacement is proportional to the signed area enclosed by the loop. (b) Projection of a finite trajectory segment. The shaded region denotes the signed area swept relative to the origin; the faint curve indicates the continuation of the trajectory.}
    \label{fig:area-to-propulsion}
\end{figure}

The geometric interpretation is illustrated in
figure~\ref{fig:area-to-propulsion}. For a periodic trajectory, the
projection onto an adjacent mode plane forms a closed loop
(figure~\ref{fig:area-to-propulsion}(a)). The swept area then reduces to the
signed area enclosed by the loop, so that the displacement accumulated
during one cycle is proportional to this enclosed area. For a finite
trajectory segment
(figure~\ref{fig:area-to-propulsion}(b)), the projection is generally open.
Joining every point of the projected curve to the origin generates a
ruled surface whose signed area is precisely
$A_{\ell,\ell+1}(t)$. In this case, the displacement is determined by the
continuously accumulated swept area together with the endpoint
contribution $\Phi_\ell(t)-\Phi_\ell(0)$.

The displacement accumulated during the $n$th forcing period defines the
cycle-averaged propulsion velocity
\begin{equation}
\begin{aligned}
\bar U_n
& =
\frac1T
\int_{nT}^{(n+1)T}
U(t)\,dt\\
&=
\frac{Z((n+1)T)-Z(nT)}{T},
\end{aligned}
\label{eq:cycle_averaged_U}
\end{equation}
which forms a discrete time series sampled once per forcing cycle.

A cumulative average over an observation time $\hat{T}$ is then given by
\begin{equation}
\bar U(\hat{T})=\frac{\Delta Z(\hat{T})}{\hat{T}},
\end{equation}
and, whenever the limit exists, defines the long-time mean propulsion
\begin{equation}
\langle U\rangle
=
\lim_{\hat{T}\rightarrow\infty}
\bar U(\hat{T}).
\end{equation}

Equation~\eqref{eq:ZPropulsion} has the same mathematical structure as
the geometric formulation of low-Reynolds-number swimming introduced by
Shapere and Wilczek~\cite{ShapereWilczek1987, ShapereWilczek1989}, in which the displacement is represented as a line
integral of a gauge potential $\mathcal A$ in shape space. In the present case,
\[
\mathcal A
=
\sum_{\ell\ge2}
\mathcal C_\ell
\left(
q_\ell\,dq_{\ell+1}
-
q_{\ell+1}\,dq_\ell
\right),
\]
while the symmetric contribution corresponds to the exact differential
$d\Phi$. For closed trajectories, the endpoint terms vanish, leaving
only the gauge-invariant area contribution.

The propulsion problem therefore separates naturally into two stages.
The active forcing first generates trajectories on the constrained shape
manifold through the nonlinear mode dynamics. The propulsion law then
maps the geometric properties of these trajectories onto rigid-body
translation.

\subsection{Asymptotic regimes of propulsion dynamics}
\label{subsec:asymptotic_regimes}

Several characteristic regimes of the coupled shape and propulsion
dynamics can be understood directly from asymptotic limits of the
projected mode equation~\eqref{eq:projectedDynamics} together with the
propulsion law~\eqref{eq:ZPropulsion}. The following analysis
characterizes the generic behavior of trajectories whose projections
onto adjacent mode planes sweep a nonzero signed area.
 
In the absence of forcing, the projected dynamics reduces to a passive
relaxation flow on the constraint sphere, with the Helfrich energy as a
Lyapunov function, as shown in appendix~\ref{app:Lyapunov}. The passive
dynamics minimizes the elastic energy by concentrating the excess area
into the softest accessible mode, namely $\ell=2$. Consequently, the
quadrupolar states
\begin{equation}
q_2=\pm1,
\qquad
q_{\ell>2}=0,
\label{eq:global_stable_fp}
\end{equation}
are stable fixed points.

Denoting the strength of the forcing by $s$, $F=O(s)$, we consider the regime of weak forcing with $s\ll 1$. In this case the dynamics consists of small
excursions about these passive equilibria,
\begin{equation}
\delta q_\ell = O(s).
\end{equation}
If the projected trajectories in the adjacent mode planes enclose
nonzero signed area, the propulsion velocity follows from the quadratic
propulsion law and therefore scales generically as
\begin{equation}
\langle U\rangle = O(s^2).
\end{equation}

In the slow-driving regime, the forcing period
$T=2\pi/\omega$ is much larger than the longest intrinsic relaxation
time $\tau$ on the constraint sphere, so that $\omega\tau\ll1$. For the
constrained linear Helfrich dynamics, the slowest transverse relaxation
away from the passive $\ell=2$ branch is governed by the $\ell=3$ mode,
\begin{equation}
\tau^{-1}
=
\Gamma_3(\beta_3-\beta_2).
\end{equation}
When
\[
\omega \ll \tau^{-1},
\]
the deformation follows the instantaneous quasistatic equilibrium,
\[
\bm q(t)
=
\bm q_{\rm qs}(\omega t)
+
O(\omega\tau).
\]
Consequently,
\begin{equation}
\langle U\rangle = O(\omega),
\end{equation}
provided the projected trajectories enclose a nonzero signed area.

In the opposite limit of fast driving,
\[
\omega\gg\tau^{-1},
\]
the forcing varies on a timescale much shorter than the intrinsic
relaxation time. The deformation therefore separates into a slowly
evolving passive component and a small oscillatory correction,
\[
\bm q(t)
=
\bar{\bm q}
+
\delta\bm q(t),
\qquad
\delta\bm q
=
O(\omega^{-1}).
\]
Averaging over the fast forcing period eliminates the leading
oscillatory forcing contribution, leaving the passive Helfrich dynamics
for the slowly varying component $\bar{\bm q}$. Since the static
$\ell=2$ component does not contribute to the cycle-averaged
propulsion, the net displacement originates entirely from the
phase-lagged oscillatory response. Consequently,
\begin{equation}
\langle U\rangle
=
O(\omega^{-1}),
\end{equation}
up to additional cancellations imposed by symmetry.
Thus the mean propulsion vanishes in both the quasistatic and the high-frequency limits and is therefore expected to attain a maximum at intermediate driving frequencies, where the forcing period becomes comparable to the intrinsic relaxation time.

A final asymptotic regime is obtained for strong driving,
$s\gg1$, where the active forcing dominates passive elastic relaxation, 
and equation~\eqref{eq:projectedDynamics} reduces to
\[
\dot{\bm q}
\approx
-\hat{\bm P}(\bm q)\hat{\bm\Gamma}\bm F(t).
\]
The shape then relaxes toward the instantaneous stable branch antiparallel to $\bm F(t)$, 
while remaining on the constraint sphere.
The relaxation rate toward the stable branch scales as $O(s)$,
whereas the forcing varies on the timescale $O(\omega^{-1})$. Whenever
the alignment relaxation is much faster than the forcing period, the
shape adiabatically follows the instantaneous forcing direction.
Consequently, strong forcing enslaves the shape dynamics,
eliminates independent internal frequencies, and generically drives the
system toward synchronization with the external forcing. 

 Taken together, these asymptotic limits identify three broad regimes of the driven vesicle dynamics. 
 The first is a relaxation-dominated regime, reached either for weak forcing or sufficiently rapid driving. There, the deformation remains close to the passive Helfrich
equilibria. Weak forcing produces small oscillations about the passive
$\ell=2$ branch, whereas fast driving averages over the forcing and
leaves only small phase-lagged oscillatory corrections. 
 The second is a drive-dominated (adiabatically slaved) regime, reached either for sufficiently slow driving or sufficiently strong forcing. In this case the deformation closely follows the instantaneous
quasistatic equilibrium or forcing direction and therefore evolves
predominantly under the control of the external drive.
Between these asymptotic limits lies an intermediate regime in which
the forcing amplitude, forcing frequency, and intrinsic relaxation act
on comparable scales. It is in this regime that the richest nonlinear dynamics emerges.

The asymptotic regimes discussed above can be related to experimentally
relevant time scales in active membranes and cortical systems. The
intrinsic bending relaxation time depends strongly on vesicle size,
viscosity contrast, and mode number. For quasi-spherical vesicles, the
relaxation times of the lowest modes scale as
\[
\tau_\ell
\sim
\frac{\eta a^3}{\kappa\,\ell^3},
\]
up to geometric and viscosity-ratio prefactors. This corresponds to
relaxation times ranging from milliseconds for micron-sized lipid
vesicles to seconds or longer for giant vesicles and cell-sized
membranes.

Active membrane and cortical processes span comparable time scales,
ranging from fractions of a second to several minutes depending on the
underlying biochemical mechanisms and transport processes
\cite{Mayer2009,Bement2015,Michaud2022}. The asymptotic regimes
identified above are therefore expected to be experimentally
accessible. In particular, cell-sized active membranes are likely to
operate in the intermediate regime, where forcing and relaxation act on
comparable time scales and the richest nonlinear dynamics is expected.

\section{Reduced dynamics with two and three modes}
\label{sec:general3modes}

The projected dynamics simplifies considerably when restricted to two or three adjacent modes.
The
two-mode reduction confines the dynamics to a circle and is therefore described by a single phase variable. By
contrast, the three-mode reduction is the minimal setting that permits
genuinely two-dimensional long-time dynamics on the constraint sphere.
These reduced systems provide the foundation for the analysis of the
nonlinear dynamics in the following sections.

\subsection{Angular velocity and dynamics in mode space}

For definiteness, we consider the mode triple $(2,3,4)$, although the
construction extends directly to any ordered triple of adjacent modes.

The motion on the constraint sphere is conveniently characterized by
the angular velocity vector in mode space,
\begin{equation}
\bm\Omega
=
\bm q\times\dot{\bm q}.
\label{eq:angular_velocity_modespace}
\end{equation}
Its components are the signed areal sweep rates in the three projected
mode planes,
\begin{equation}
\dot A_{ij}
=
\frac12
\left(
q_i\dot q_j-q_j\dot q_i
\right),
\qquad
(i,j,k)\ \text{cyclic},
\label{eq:area_rates}
\end{equation}
so that
\[
\Omega_k
=
2\dot A_{ij}.
\]

The projected evolution equation (\ref{eq:projectedDynamics}) can be
cast into a rotational form by multiplying it with
$\mu=\bm q\cdot\bm Q$. This
 gives
\[
\mu\dot{\bm q}
=
\bm Q(\bm q\cdot\bm F^p)-\mu\bm F^p
=
\bm q\times(\bm Q\times\bm F^p),
\]
or equivalently,
\begin{equation}
\mu\dot{\bm q}
=
\bm q\times\bm\tau,
\qquad
\bm\tau=\bm Q\times\bm F^p.
\label{eq:tau}
\end{equation}
The constrained dynamics is therefore purely rotational on the
constraint sphere, with the state-dependent angular velocity
\[
\bm\Omega
=
-\frac{\bm\tau}{\mu}.
\]
The reaction force acts parallel to $\bm Q$ and therefore produces no
torque, so only the projected force contributes to
$\bm\tau$ (see figure~\ref{fig:oblique_projector}).
Equation~\eqref{eq:tau} provides the central geometric representation
of the constrained mode dynamics.

The mode amplitudes on the unit sphere are conveniently parametrized by
the spherical coordinates
$
q_2=\cos\psi,\quad
q_3=\sin\psi\cos\chi,\quad
q_4=\sin\psi\sin\chi.
$
The corresponding tangent basis is
\[
\bm e_\psi
=
\frac{\partial\bm q}{\partial\psi},
\qquad
\sin\psi\,\bm e_\chi
=
\frac{\partial\bm q}{\partial\chi},
\]
so that the velocity decomposes as
\[
\dot{\bm q}
=
\dot\psi\,\bm e_\psi
+
\sin\psi\,\dot\chi\,\bm e_\chi.
\]
Projecting equation~\eqref{eq:tau} onto the tangent directions yields
\begin{equation}
\mu\dot\psi=-\tau_\chi,
\qquad
\mu\sin\psi\,\dot\chi=\tau_\psi,
\end{equation}
where
\[
\tau_\psi=\bm\tau\cdot\bm e_\psi,
\qquad
\tau_\chi=\bm\tau\cdot\bm e_\chi.
\]
Explicit expressions for the torque components are given in
appendix~\ref{eq:3modeappendix}.

In the numerical integrations, we evolve continuous lifted
representatives of the angular variables rather than reducing them
modulo $2\pi$.

For two adjacent modes, the rotational dynamics reduces to a single
degree of freedom. Setting $q_4=0$ confines the motion to the great
circle $\chi=0$, and the torque equation becomes
\[
\mu\,\dot\psi
=
Q_2F^p_3-Q_3F^p_2.
\]

Inserting $\bm F^p=\hat{\bm \beta}\bm Q+\hat{\bm \Gamma}\bm F$, this becomes
\begin{align}
\mu\,\dot\psi
= &
-\Gamma_2\Gamma_3
\Big[
(\beta_3-\beta_2)\sin\psi\cos\psi +\\\nonumber
& +\cos\psi\,F_3
-\sin\psi\,F_2
\Big].
\label{eq:psi_2mode}
\end{align}

with 
$
\mu
=\Gamma_2\cos^2\psi+\Gamma_3\sin^2\psi.
$
Thus the two-mode dynamics is governed by a single nonlinear equation
for the angle $\psi$, whose intrinsic part is controlled by the anisotropy
difference $\beta_3-\beta_2$ and whose driven part is the tangential
projection of the applied force in the $(q_2,q_3)$ plane.
\subsection{Propulsion velocity}

The propulsion law acquires a particularly simple form when expressed
in terms of the angular velocity in mode space,
\begin{equation}
U(t)
=
\bm {\mathcal C}\cdot\bm\Omega
+
\frac{d\Phi}{dt},
\label{eq:U_area_general}
\end{equation}
where
\[
\bm{\mathcal C}=(\mathcal C_3,0,\mathcal C_2)^t.
\]
The vector $\bm{\mathcal C}$ specifies how angular motion in mode space is
converted into propulsion and may therefore be interpreted as a
deformation--translation coupling vector. Since the total derivative
does not contribute to the long-time average,
\[
\langle U\rangle
=
\bm{\mathcal C}\cdot\langle\bm\Omega\rangle.
\]

For a periodically driven system with forcing period
$T=2\pi/\omega$, it is convenient to normalize the accumulated angular
rotation per forcing cycle and introduce the dimensionless,
plane-wise rotation numbers
\begin{equation}
\rho_k
=
\lim_{n\to\infty}
\frac{1}{2\pi n}
\int_0^{nT}\Omega_k(t)\,dt
=
\frac{T}{2\pi}\,\overline{\Omega}_k.
\label{eq:plane_rotation_number_periodic}
\end{equation}
The mean propulsion velocity then becomes
\[
\langle U\rangle
=
\omega\,\bm{\mathcal C}\cdot\bm\rho,
\]
establishing a direct link between the nonlinear dynamics in mode space
and the resulting propulsion.

%%%%%%%%%%%%%%%%%%%%%%%%%%%%%%%%%%%%%%%%%%%%%%%%%%%%
\section{Two--mode dynamics}
\label{sec:twomode}
%%%%%%%%%%%%%%%%%%%%%%%%%%%%%%%%%%%%%%%%%%%%%%%%%%%%%

We begin with the simplest mode truncation capable of generating
propulsion by retaining only the adjacent modes $\ell=2$ and $\ell=3$.
These are the lowest nontrivial deformation modes, while higher modes
relax on progressively shorter time scales. The resulting
one-dimensional dynamics provides a transparent setting in which to
relate synchronization and phase locking directly to propulsion.

Throughout this section, we consider harmonic forcing of the form
\begin{equation}
F_\ell(t)=s\alpha_\ell\cos(\omega t + \delta_\ell)
\end{equation}
and choose $\alpha_2=1$, $\alpha_3=\alpha$, $\delta_2=0$ and $\delta_3=\delta$ without loss of generality.

\subsection{Phase reduction and propulsion}

The area constraint confines the two-mode dynamics to the unit circle,
which we parameterize by the phase angle $\psi$,
\[
q_2=\cos\psi,
\qquad
q_3=\sin\psi.
\]
Since the physical state depends only on $\tilde\psi=\psi\bmod2\pi$, we evolve a
continuous lifted representative $\psi$ of the phase, making the rotation
number well defined. The only non-vanishing component of the angular
velocity is
\[
\Omega_4
=
q_2\dot q_3-q_3\dot q_2
=
\dot\psi.
\]
Consequently, the propulsion law reduces to
\begin{equation}
U
=
\dot\psi
\bigl(\mathcal C_2+B_2\cos2\psi\bigr).
\label{eq:U_two_mode_compact_section}
\end{equation}
Averaging over many forcing periods gives
\begin{equation}
\langle U\rangle
=
\omega\, \mathcal C_2\,\rho,
\label{eq:U_rotation_relation_section}
\end{equation}
where
\begin{equation}
\rho
=
\lim_{n\to\infty}
\frac{\psi_n-\psi_0}{2\pi n},
\qquad
\psi_n=\psi(nT),
\label{eq:2mode_rotation_number}
\end{equation}
follows directly from the general definition
equation~\eqref{eq:plane_rotation_number_periodic}  with $\rho=\rho_4$, since
$\Omega_4=\dot\psi$. The propulsion problem is therefore reduced to determining 
the rotation number of the stroboscopic dynamics.

\subsection{Locked and running states}
Without forcing ($s=0$), the reduced phase dynamics possesses two
stable fixed points at
\[
\tilde\psi=0,\pi,
\]
corresponding to the pure $q_2$ states $(\pm1,0)$, and two
unstable fixed points at
\[
\tilde\psi=\pi/2,3\pi/2,
\]
corresponding to the pure $q_3$ states $(0,\pm1)$.
For sufficiently weak forcing, this structure persists, giving rise to
two stable librating states. The phase remains confined to the vicinity
of one of the stable equilibria. Consequently,
$
\rho=0,
$
and the mean propulsion vanishes despite the ongoing shape
oscillations.

\begin{figure}[htbp]
    \centering
    \begin{subfigure}[t]{0.49\columnwidth}
      \includegraphics[width=\linewidth]{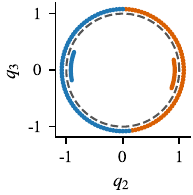}
      \caption{}
    \end{subfigure}
    \begin{subfigure}[t]{0.49\columnwidth}
      \includegraphics[width=\linewidth]{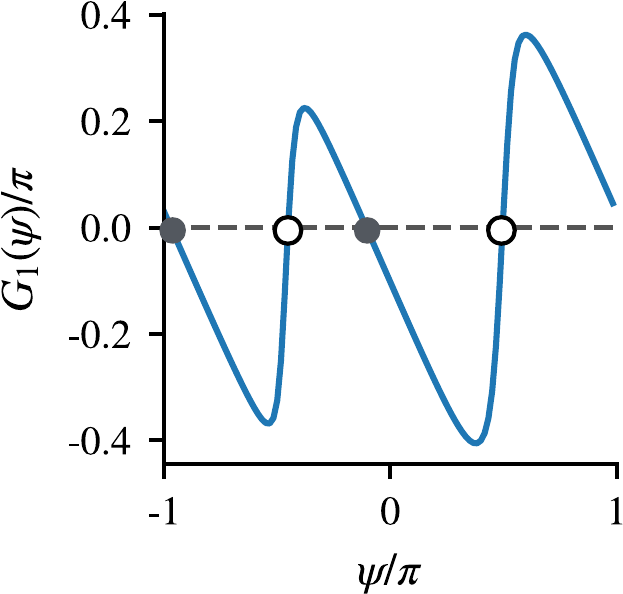}
      \caption{}
    \end{subfigure}

\caption{
Librating states in the two-mode dynamics.
(a) Basins of attraction and attractors of the two symmetry-related
librating states. The outer arcs indicate the basins of attraction,
while the inner arcs show the corresponding attractors.
(b) Poincar\'e increment $G(\psi)$.
Stable and unstable fixed points are indicated by filled and open
symbols, respectively. The stable fixed points correspond to the two
librating attractors shown in panel~(a).
}
    \label{fig:bistable_panel}
\end{figure}

The transition from librating to running motion is conveniently analyzed
using two complementary diagnostics: the stroboscopic map and the lifted
phase evolution~\cite{Strogatz2014}. We first introduce the stroboscopic map,
\begin{equation}
\psi_{n+1}=P(\psi_n),
\label{eq:poincare_map}
\end{equation}
obtained by sampling the phase once every forcing period
$T=2\pi/\omega$.
The map identifies periodic states and visualizes the phase-space
dynamics, while the lifted phase distinguishes bounded oscillations from
sustained phase rotation. Stable fixed points of $P$ correspond to phase-locked
librating states. A representative example is shown in
figure~\ref{fig:bistable_panel}. Panel (a) displays the two librating
attractors together with their basins of attraction on the unit circle,
while panel (b) shows the corresponding Poincaré increment
\begin{equation}
G(\psi)=P(\psi)-\psi,
\end{equation}
whose stable zeros coincide with the attracting fixed points.

\begin{figure}
    \centering
    \compactpanelgraphic[width=0.49\columnwidth]{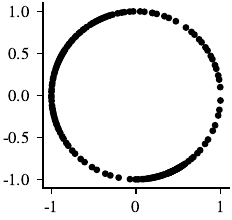}{a}\hfill
    \compactpanelgraphic[width=0.49\columnwidth]{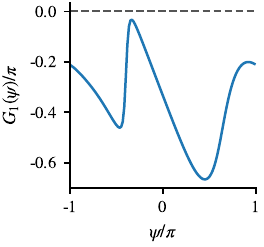}{b}
    \par
    \compactpanelgraphic[width=0.49\columnwidth]{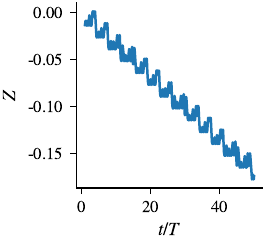}{c}\hfill
    \compactpanelgraphic[width=0.49\columnwidth]{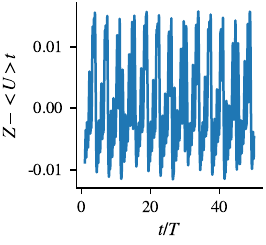}{d}
    \caption{
Quasiperiodic running state in the two-mode dynamics.
(a) Stroboscopic dynamics on the unit circle. 
(b) Poincar\'e increment $G(\psi)$. 
(c) Displacement $Z(t)$ showing a finite mean propulsion.
(d) Detrended displacement $Z-\langle U\rangle t$.
}
    \label{fig:quasiperiodic_running}
\end{figure}

\begin{figure}
    \centering
    \compactpanelgraphic[width=0.49\columnwidth]{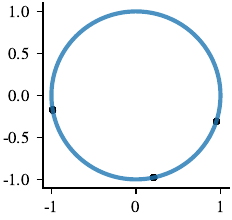}{a}\hfill
    \compactpanelgraphic[width=0.49\columnwidth]{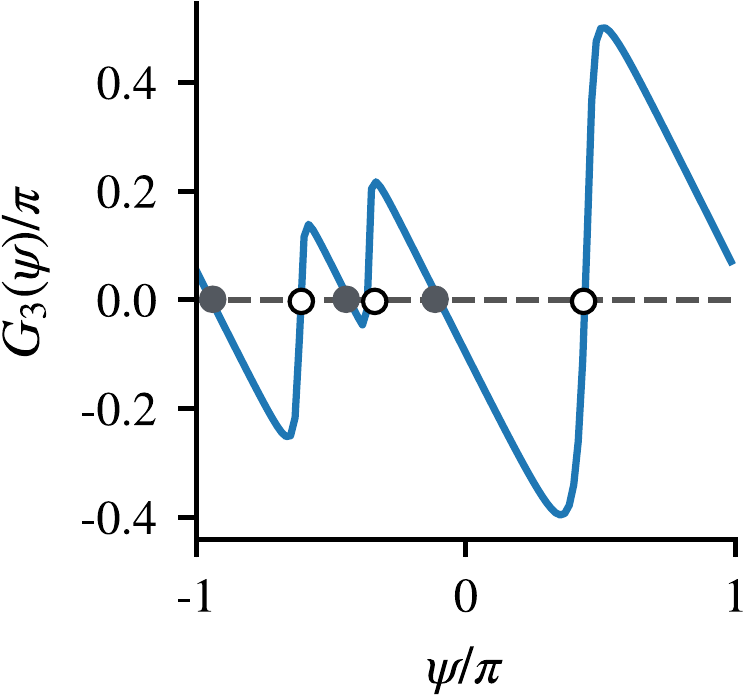}{b}
    \par
    \compactpanelgraphic[width=0.49\columnwidth]{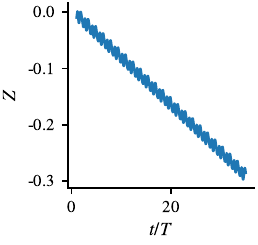}{c}\hfill
    \compactpanelgraphic[width=0.49\columnwidth]{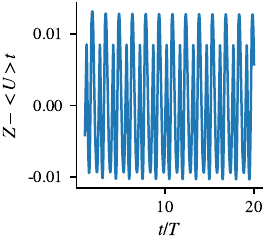}{d}
  \caption{
Phase-locked running state in the two-mode dynamics.
(a) Stroboscopic dynamics on the unit circle.
(b) Poincar\'e increment $G(\psi)$. Stable and unstable zeros correspond to the attracting periodic orbit and its associated unstable counterpart.
(c) Displacement $Z(t)$ showing a finite mean propulsion.
(d) Detrended displacement $Z-\langle U\rangle t$. 
}
    \label{fig:cyclic_running}
\end{figure}

Running motion sets in when the stable fixed points of $P(\psi)$
disappear. The lifted phase then grows without bound, corresponding to
a continuous rotation of the deformation state around the constraint
circle. Running states may be either phase locked or quasiperiodic. For 
a rational rotation number,
$
\rho=p/q,
$
the dynamics is phase locked and satisfies
\[
P^q(\psi)=\psi+2\pi p.
\]
For irrational $\rho$, the phase evolves quasiperiodically and
densely fills the circle~\cite{KatokHasselblatt1995}.
Representative examples of both regimes are shown in
figures~\ref{fig:quasiperiodic_running} and~\ref{fig:cyclic_running}. The lower panels display the
corresponding displacement $Z(t)$ together with the detrended
displacement
\[
Z(t)-\langle U\rangle t.
\]
After subtracting the mean translation, the remaining oscillatory
component is periodic for phase-locked running, whereas it remains
bounded but never repeats in the quasiperiodic regime.

\subsection{Symmetry and unlocking transition}
\label{subsect:symmetry}

\begin{figure*}
    \centering
    \begin{subfigure}[t]{0.3\linewidth}
        \centering
        \includegraphics[width=0.9\linewidth]{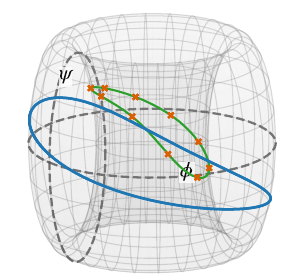} 
        \caption{Symmetry-related attractors} \label{fig:timing1}
    \end{subfigure}
    \hfill
    \begin{subfigure}[t]{0.3\linewidth}
        \centering
        \includegraphics[width=0.9\linewidth]{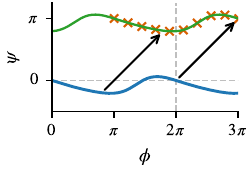} 
        \caption{Attractors of (a) in the $(\psi,\varphi)$ chart} \label{fig:timing2}
    \end{subfigure}
    \hfill
    \begin{subfigure}[t]{0.3\linewidth}
        \centering
        \includegraphics[width=0.85\linewidth]{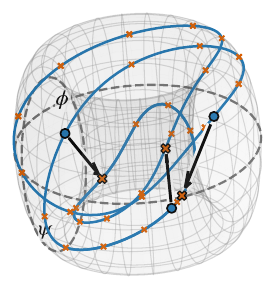} 
        \caption{Symmetric 3-cycle attractor} \label{fig:timing3}
    \end{subfigure}

 \caption{
Action of the time-glide symmetry on locked and running attractors.
Trajectories are shown in the extended phase space $(\psi,\varphi)$.
(a) In the bistable regime, the two phase-locked attractors are exchanged by the half-period shift
$(\psi,\varphi)\mapsto(\psi+\pi,\varphi+\pi)$.
The markers indicate the symmetry image of one attractor on the other.
(b) The same attractors in the unfolded $(\psi,\varphi)$ representation, where the symmetry acts as a translation by $(\pi,\pi)$.
(c) In the running regime, the symmetry acts within a single phase-locked orbit. Circle and cross markers denote symmetry-related points on the same attractor, while the arrows indicate the induced mapping.
}
\label{fig:extended_torus_symmetry}
\end{figure*}

The time-glide symmetry,
\[
(\psi,\varphi)\mapsto(\psi+\pi,\varphi+\pi),
\]
introduced in section~\ref{subsec:time_glide_symmetry}, has qualitatively
different consequences in the librating and running regimes. In the
bistable regime, it exchanges the two coexisting attractors, as
illustrated in figure~\ref{fig:extended_torus_symmetry}(a) and (b).
In the intrinsic $(\psi,\varphi)$ representation of panel~(b), the symmetry
is simply a translation by $(\pi,\pi)$. Its action is less apparent in
the embedded representation of panel~(a), owing to the nonlinear
embedding of the constraint manifold.
The symmetry requires the two attractors to disappear simultaneously at the unlocking threshold
$s=s^*(\omega,\alpha,\delta)$. Beyond this transition, all parameter
regimes investigated exhibit a single running attractor. The half-period
shift then acts within this attractor, which is invariant under the
transformation. An example of such a self-symmetric orbit is shown in
figure~\ref{fig:extended_torus_symmetry}(c), where symmetry-related
points on the same trajectory are connected by the induced mapping.

\begin{figure}
    \centering
    \includegraphics[width=\linewidth]{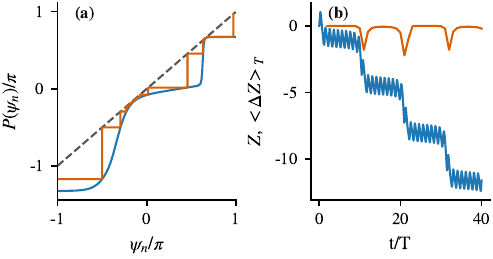}
\caption{
Intermittent dynamics near the unlocking transition.
(a) Cobweb construction of the Poincar\'e map close to the unlocking threshold.
(b) Corresponding displacement $Z(t)$ (blue) and displacement per forcing cycle $\Delta Z(T)$ (orange). 
}
    \label{fig:intermittency}
\end{figure}

The nature of the unlocking transition becomes apparent when the
threshold $s=s^*$ is approached from above. Close to onset, the
trajectory exhibits critical slowing down as it passes the ghosts of
the annihilated fixed-point pair. Long intervals of nearly locked
motion are interrupted by rapid phase slips that advance $\psi$ by
$2\pi$. Each phase slip produces a burst of propulsion, giving rise to
the intermittent displacement shown in
figure~\ref{fig:intermittency}.

This behaviour is the characteristic signature of a saddle-node
bifurcation on an invariant circle (SNIC)~\cite{Strogatz2014, Kuznetsov1998}. Near the bifurcation, the
rotation number obeys the universal square-root scaling
\begin{equation}
\rho\sim (s-s^*)^{1/2},
\qquad
\langle U\rangle\sim (s-s^*)^{1/2},
\end{equation}
as derived in appendix~\ref{app:snic_map}. Consequently, the mean
propulsion sets in continuously at the unlocking threshold.

\subsection{Phase locking and transport plateaus}

\begin{figure}
    \centering
    \begin{subfigure}[t]{\columnwidth}
        \centering
        \includegraphics[width=\linewidth]{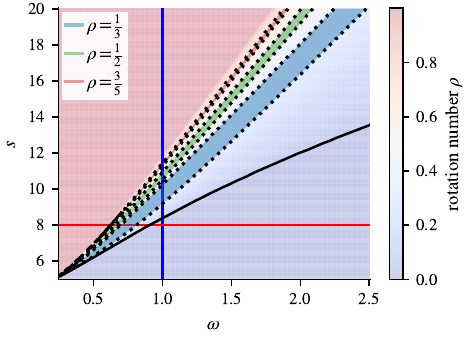} 
        \caption{} \label{fig:Arnold}
    \end{subfigure}

    \begin{subfigure}[t]{0.49\columnwidth}
         \includegraphics[width=\columnwidth]{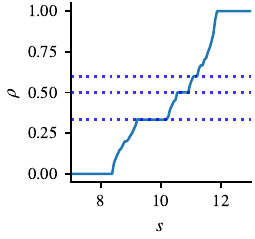} 
         \caption{} \label{fig:two_mode_scan_panel_a}
     \end{subfigure}
      \begin{subfigure}[t]{0.49\columnwidth}
           \includegraphics[width=\columnwidth]{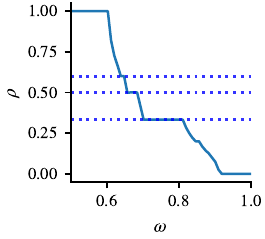} 
         \caption{} \label{fig:two_mode_scan_panel_b}
     \end{subfigure}
   \caption{
Mode-locking phase diagram of the two-mode dynamics.
(a) Rotation number $\rho(\omega,s)$ as a function of driving frequency
and forcing amplitude for $\alpha=5/7$ and $\delta=0.6\pi$.
The solid curve marks the unlocking boundary ($\rho=0$).
Selected Arnold tongues are labeled by their rational rotation numbers,
and the dotted curves indicate their boundaries.
The horizontal line corresponds to the forcing-amplitude sweep shown in
the left panel of (b), while the vertical line corresponds to the
driving-frequency sweep shown in the right panel.
(b) Rotation number as a function of forcing amplitude (left) and
driving frequency (right) along the cuts indicated in panel~(a). Flat
segments correspond to phase-locked states.
}
\label{fig:s-omega-scan}
\end{figure}

Beyond the primary unlocking transition, the running regime exhibits
additional regions of rational phase locking.
Figure~\ref{fig:s-omega-scan}(a) shows the rotation number
$\rho(s,\omega)$ in the two-parameter plane for fixed
$\alpha=5/7$ and $\delta=0.6\pi$. The regions with rational values of
$\rho$ form the familiar Arnold tongues of periodically driven
systems~\cite{guckenheimer_holmes, Pikovsky2001}.

The horizontal and vertical cuts shown in
figure~\ref{fig:s-omega-scan}(b) and (c) illustrate how the rotation number
varies along one-parameter sweeps in forcing amplitude and driving
frequency. Within each Arnold tongue, the rotation number remains
locked to the rational value $\rho=p/q$. Between the tongues, the
motion is quasiperiodic and the rotation number varies continuously
with the control parameters.

Since the mean propulsion satisfies
$\langle U\rangle\propto\omega\rho$, synchronization is reflected
directly in the transport. For fixed driving frequency, the mean
propulsion exhibits plateaus as the forcing amplitude is varied.
Conversely, for fixed forcing amplitude, it forms piecewise linear
branches as a function of the driving frequency, separated by kinks at
the boundaries of the Arnold tongues. Thus, the synchronization
structure of the underlying phase dynamics is directly encoded in the
propulsion velocity.

%%%%%%%%%%%%%%%%%%%%%%%%%%%%%%%%%%%%%%%%%%%%%%%%%%%%%%%%%%%%%%%%%%%%%%%%%%%%%%%%%%%%%%%%%
\section{Three--mode dynamics}
\label{sec:threemode}
%%%%%%%%%%%%%%%%%%%%%%%%%%%%%%%%%%%%%%%%%%%%%%%%%%%%%%%%%%%%%%%%%%%%%%%%%%%%%%%%%%%%%%%%%

The two-mode reduction is completely described by a single phase
variable, allowing both the deformation dynamics and the resulting
propulsion to be analyzed using one-dimensional dynamical-systems
techniques. The three-mode system is the minimal extension beyond this
setting: its second independent deformation degree of freedom permits
genuinely higher-dimensional dynamics. Our goal is therefore twofold: to
identify the new classes of attractors that emerge in the reduced
deformation space and to determine how this richer dynamics is
reflected in experimentally accessible propulsion observables.

In analogy to the two-mode system, we study harmonic drives of the form 

\begin{equation}
    F_\ell(t) = s\alpha_\ell \cos(\omega t + \delta_\ell),
\end{equation}
with $\alpha_2=1$ and $\delta_2=0$. Throughout this section, all
numerical results use
$\alpha_2=\alpha_3=\alpha_4=1$, $\delta_3=1.55$, and $\delta_4=1.0$.

\subsection{Passive equilibria and weak forcing}

As in the two-mode system, we begin by considering the passive dynamics $(s=0)$. Its fixed points provide the reference states for the weakly forced dynamics.

Introducing the configuration-dependent weighted average of relaxation rates,
\begin{equation}
\bar\beta(\bm q)
=
\frac{\sum_{i=2}^4 \beta_i\Gamma_i q_i^2}
{\sum_{i=2}^4 \Gamma_i q_i^2},
\end{equation}
equation~\eqref{eq:projectedDynamics} can be written componentwise as
\begin{equation}
\dot q_i
=
\Gamma_i q_i\bigl(\bar\beta-\beta_i\bigr),
\qquad i=2,3,4.
\label{eq:3mode_component_dynamics}
\end{equation}

Because $\bar\beta$ is a weighted average of the $\beta_i$, the condition
$\bar\beta=\beta_i$ can only be satisfied if all remaining components
vanish. For distinct $\beta_i$, the only fixed points are therefore the
six pure-mode states
\[
\bm q^\ast=\pm\bm e_2,\,
\pm\bm e_3,\,
\pm\bm e_4.
\]

As shown in section~\ref{subsec:asymptotic_regimes}, the Lyapunov function already establishes
that $\pm\bm e_2$ are the globally attracting fixed points (see equation~\eqref{eq:global_stable_fp}). To determine
the stability of the remaining pure-mode states, and to recover the local
stability of all fixed points in a unified way, we linearize the dynamics
about each $\bm q^\ast$. Writing
\[
\bm q=\bm e_k+\bm\xi,
\qquad
\bm\xi\cdot\bm e_k=0,
\]
the linearized dynamics in the tangent plane becomes
\begin{equation}
\dot\xi_j
=
\Gamma_j(\beta_k-\beta_j)\xi_j,
\qquad
j\neq k.
\end{equation}

The stability is therefore determined by the differences $\beta_k-\beta_j$.
Since $\beta_2<\beta_3<\beta_4$, the states $\pm\bm e_2$ are stable
nodes, $\pm\bm e_3$ are saddle points, and $\pm\bm e_4$ are unstable
nodes.

In the two-mode reduction $(q_2,q_3)$, the pure $q_3$ states appear as
unstable fixed points. In the three-mode system, these fixed points are
resolved into saddle points whose stable manifolds separate the basins
of attraction of the two stable equilibria $\pm\bm e_2$.

For sufficiently weak forcing, the dynamics remains close to one of the stable passive equilibria. The forced motion can therefore be obtained by linearizing about, for example, 
$\bm q=\bm e_2$. Since the unit-sphere constraint eliminates radial motion,
the dynamics is
confined to the tangent plane spanned by the transverse modes $q_3$ and
$q_4$.
To leading order one obtains two independently driven relaxators,
\begin{equation}
\begin{aligned}
\dot q_3+\lambda_3 q_3&=-\Gamma_3F_3(t),\\
\dot q_4+\lambda_4 q_4&=-\Gamma_4F_4(t),
\end{aligned}
\end{equation}
with relaxation rates
\begin{equation}
\lambda_3=\Gamma_3(\beta_3-\beta_2),\qquad
\lambda_4=\Gamma_4(\beta_4-\beta_2).
\end{equation}

For harmonic forcing,
the two tangent modes respond with phase lags
\[
\phi_i=\arctan(\omega/\lambda_i).
\]
Substituting the linear response into the propulsion law gives
\begin{equation}
\begin{aligned}
\langle U\rangle
={}&
\frac{C_{3}\omega\Gamma_3\Gamma_4\,s^2}
{\sqrt{\lambda_3^2+\omega^2}\sqrt{\lambda_4^2+\omega^2}}
\\
&\times\sin\!\Big[(\delta_4-\delta_3)+(\phi_3-\phi_4)\Big]
+\mathcal O(s^3).
\end{aligned}
\label{eq:Ubar_weak_3mode}
\end{equation} 
The dependence on the driving strength $s$ and frequency $\omega$ is
consistent with the weak-forcing and high- and low-frequency asymptotic
regimes discussed in section~\ref{subsec:asymptotic_regimes}.

Equation~\eqref{eq:Ubar_weak_3mode} shows that the linear shape response
already produces nonzero mean propulsion whenever the two transverse
deformation modes acquire a relative phase lag. This lag may be imposed
directly through the forcing phases $\delta_3,\delta_4$, or generated
internally by the unequal relaxation rates $\lambda_3$ and
$\lambda_4$.

This contrasts sharply with the two-mode dynamics. There, weak forcing
produces only librating trajectories on the constraint circle, so the
rotation number vanishes and no mean propulsion occurs. In the
three-mode system, the additional transverse degree of freedom permits
closed trajectories with nonzero oriented area already in the
weak-forcing regime.

\begin{figure}
    \centering
    \begin{subfigure}[t]{0.45\linewidth}
        \centering
        \includegraphics[width=\linewidth]{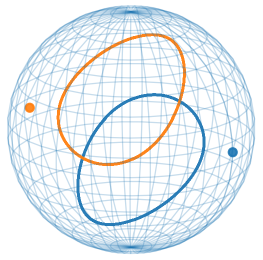} 
        \caption{Synchronized cycles} \label{fig:3_modes_weak_forcing}
    \end{subfigure}
    \hfill
    \begin{subfigure}[t]{0.45\linewidth}
        \centering
        \includegraphics[width=\linewidth]{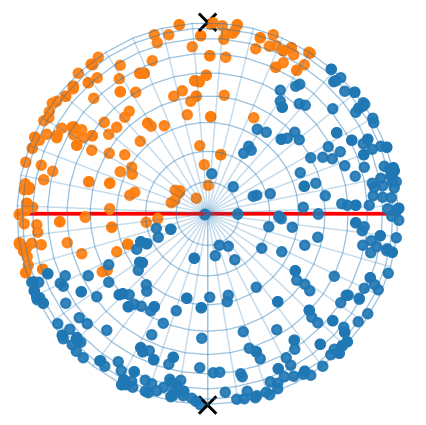} 
        \caption{Basins of attraction} \label{fig:3_modes_basin}
    \end{subfigure}
    \caption{Coexisting synchronized attractors near the synchronization threshold $(s=7.53,\, \omega=1.48)$. (a) The two stable phase-locked cycles related by the discrete spatiotemporal symmetry discussed in section~\ref{subsec:time_glide_symmetry}. (b) Pole-to-pole view of the basins of attraction on the unit sphere of shape space. Initial conditions converge to one of the two synchronized cycles.
    }
    \label{fig:basin}
\end{figure}

Figure~\ref{fig:basin} illustrates the weakly forced bistable regime
immediately below the primary unlocking transition. The two stable
periodic states remain surrounded by finite basins of attraction,
showing that bistability persists up to the transition rather than
disappearing gradually. This regime provides the reference state
against which the more complex dynamics at larger forcing amplitudes
will be compared.

\subsection{Intermediate forcing: invariant tori}

As the forcing amplitude is increased beyond the weak-forcing regime,
the synchronized oscillations about the two stable equilibria grow in
size and eventually disappear at a critical amplitude
$s^*(\omega)$. Because the two attractors are related by the discrete
symmetry (\ref{eq:symmetry}), they disappear simultaneously and are
replaced by a single global attractor. At the transition, the basins of
attraction of the synchronized cycles remain finite.

\begin{figure}
    \centering
    \begin{subfigure}[t]{0.45\linewidth}
        \centering
        \includegraphics[width=\linewidth]{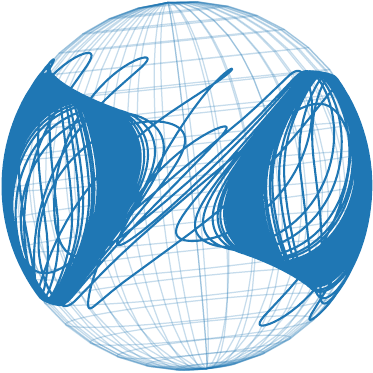} 
        \caption{} \label{fig:3_modes_torus_traj}
    \end{subfigure}
    \hfill
    \begin{subfigure}[t]{0.45\linewidth}
        \includegraphics[width=\linewidth]{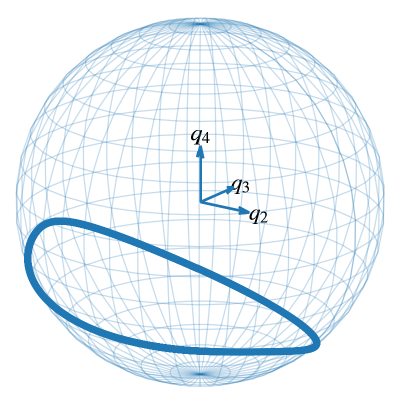} 
        \caption{} \label{fig:3_modes_torus_strobe}
    \end{subfigure}
    \caption{Dynamics immediately above the synchronization transition $(s=7.537,\,\omega=1.48)$, where the attractor is an invariant torus. (a) Continuous trajectory on the attractor. (b) Stroboscopic section revealing the invariant curve associated with the quasiperiodic motion.}
    \label{fig:torus}
\end{figure}

The transition is illustrated in
figures~\ref{fig:basin} and~\ref{fig:torus}.
Figure~\ref{fig:basin} shows the two synchronized cycles immediately
before they lose stability, whereas
figure~\ref{fig:torus} shows the trajectory on the global attractor
immediately beyond the transition. Rather than approaching a periodic
orbit, the trajectory explores an extended region of the constraint
sphere.

The geometry of this attractor is most naturally understood in the
extended phase space obtained by adjoining the forcing phase
$
\varphi=\omega t \bmod 2\pi.
$
As in the two-mode analysis, the periodically driven system is thereby
rendered autonomous.
The resulting phase space is the
three-dimensional manifold
$
S^2\times S^1,
$
formed by the product of the shape sphere and the periodic forcing
phase~\cite{guckenheimer_holmes}. The resulting global attractor is a
two-dimensional invariant torus: once a trajectory reaches this
surface, it remains on it under the subsequent evolution.

The stroboscopic dynamics is obtained by recording the trajectory once
per forcing cycle, that is, by intersecting the invariant torus with the
Poincaré section $\varphi=0$. As shown in
figure~\ref{fig:torus}(b), the intersection is a closed
invariant curve rather than a finite set of periodic points.
Successive stroboscopic points do not, in general, repeat, but instead
advance continuously along the invariant curve.

Close to the transition, the toroidal motion is still strongly
influenced by the former bistable structure. Although the two
attractors have disappeared, the trajectory spends long intervals in
the regions that previously formed the centers of the two basins shown
in figure~\ref{fig:basin}(b), and only occasionally drifts between
them. The remnants of the basin geometry therefore continue to organize
the dynamics well beyond the loss of bistability.

The invariant torus encountered here should not be confused with the
torus appearing in the two-mode reduction. The latter is a kinematic
consequence of the chosen coordinates, whereas the torus of the
three-mode system is an invariant set selected by the nonlinear flow in
the extended phase space.
The distinction is reflected in the corresponding stroboscopic
dynamics. In the two-mode system, the invariant curve coincides with the
circle parametrized by the global phase variable $\psi$, so that the
dynamics is completely described by the evolution of this single
coordinate. In the three-mode system, the invariant curve is instead
determined by the dynamics and is not one of the coordinate curves of
the sphere. Although it may be parametrized by an intrinsic phase
defined along the curve, this coordinate is not known a priori but must
be reconstructed from the attractor itself. Moreover, the continuous
trajectory evolves on the associated invariant torus, of which the
invariant curve is merely the stroboscopic section. The recurrent
dynamics therefore no longer admits a prescribed global phase
description, which motivates the recurrence-based analysis introduced in the
next subsection.

\subsection{Recurrence diagnostics of torus dynamics}

Since the invariant curve is not parametrized by a distinguished global
phase coordinate, recurrent motion can no longer be characterized by a
one-dimensional phase map as in the two-mode reduction. We therefore
introduce a diagnostic that operates directly on the sequence of
stroboscopic states and detects both exact phase locking and approximate
recurrences without requiring an explicit phase coordinate on the
invariant set. Following recurrence-based approaches
\cite{Eckmann1987,Marwan2007}, we define the $k$-step recurrence distance
for every positive integer $k$ as
\begin{equation}
D_k
:=
\max_{n\in\mathcal I}
\left\|
\bm q_{n+k}-\bm q_n
\right\|,
\label{eq:recurrence_distance}
\end{equation}
where $\mathcal I$ denotes a sufficiently long segment of the
stroboscopic sequence after discarding transients. Here
$\|\cdot\|$ is the Euclidean norm in mode space.
Thus, $D_k$ measures the largest deviation from exact $k$-step
recurrence. The use of the maximum requires recurrence over the entire
trajectory rather than merely on average. Consequently, $D_k=0$ only if
the stroboscopic dynamics is periodic with a period that divides $k$.
In numerical calculations, $D_k$ is evaluated over a finite trajectory
of length $N$. Small values of $D_k$ indicate approximate recurrence,
whereas $D_k=0$, up to numerical tolerance, signals exact phase
locking.

A state phase locked to the forcing period corresponds to a single fixed
point of the stroboscopic dynamics. Accordingly,
$
D_1=0,
$
indicating locking directly to the forcing period.

More generally, if the stroboscopic dynamics is periodic with primitive
period $k_*$,
\begin{equation}
\bm q_{n+k_*}=\bm q_n
\end{equation}
for all integers $n$, so that
$
D_{k_*}=0,
$
up to numerical tolerance. The same holds for every integer multiple of
the primitive period,
$
D_{mk_*}=0,
$
and the locking order is therefore identified by the smallest positive
integer $k$ for which $D_k$ vanishes.

For quasiperiodic motion on an invariant torus, exact recurrence never
occurs, so that
$
D_k>0
$
for all $k$. Nevertheless, pronounced minima of $D_k$ may occur at
particular recurrence orders. These indicate especially close returns
of the stroboscopic trajectory and thereby quantify the emerging
resonance structure of the invariant torus.

The recurrence diagnostic applies to arbitrary mode truncations. In the two-mode reduction, however, the dynamics is already completely characterized by the iterated Poincaré map of the global phase variable discussed in section~\ref{sec:twomode}. The recurrence analysis is therefore entirely equivalent to the phase-map description and introduces no additional dynamical information. Its advantage is that the same construction carries over directly to higher-dimensional dynamics, where no globally defined phase variable exists.

\begin{figure}
    \centering
    \begin{subfigure}[t]{0.45\linewidth}
        \centering
        \includegraphics[width=\linewidth]{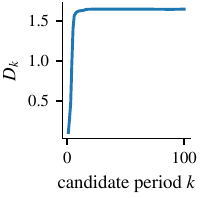} 
        \caption{} \label{fig:recurrenceDistance_s=7.537}
    \end{subfigure}
    \begin{subfigure}[t]{0.45\linewidth}
        \includegraphics[width=\linewidth]{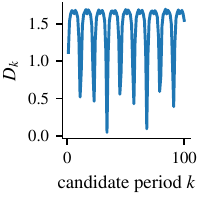} 
        \caption{} \label{fig:recurrenceDistance_s=7.75}
    \end{subfigure}
    \caption{Recurrence spectra $D_k$ for the stroboscopic dynamics. (a) Immediately above the synchronization transition $(s=7.537, \omega=1.48)$, where the newly formed invariant torus exhibits no pronounced recurrences. (b) At larger forcing amplitude $(s=7.75)$, where distinct minima develop at selected candidate periods, indicating increasingly accurate approximate recurrences. As the forcing amplitude is increased further, some of these minima eventually reach numerical zero, signalling exact resonant periodic orbits.}
    \label{fig:recurrenceDistance_traj}
\end{figure}

The evolution of the recurrence spectrum is illustrated in
figure~\ref{fig:recurrenceDistance_traj}.
Panel (a) corresponds to the parameter
values of figure~\ref{fig:torus}, immediately above the synchronization
transition. The invariant torus has only just formed, and the
recurrence spectrum is nearly featureless, exhibiting no pronounced
minima over the range of recurrence orders considered.

Further inside the quasiperiodic regime, however, a hierarchy of
increasingly deep minima develops, as shown in
figure~\ref{fig:recurrenceDistance_traj}(b). Each minimum identifies a particularly close return of the stroboscopic trajectory after a fixed number of forcing cycles.

\begin{figure}
    \centering
    \includegraphics[width=0.8\columnwidth]{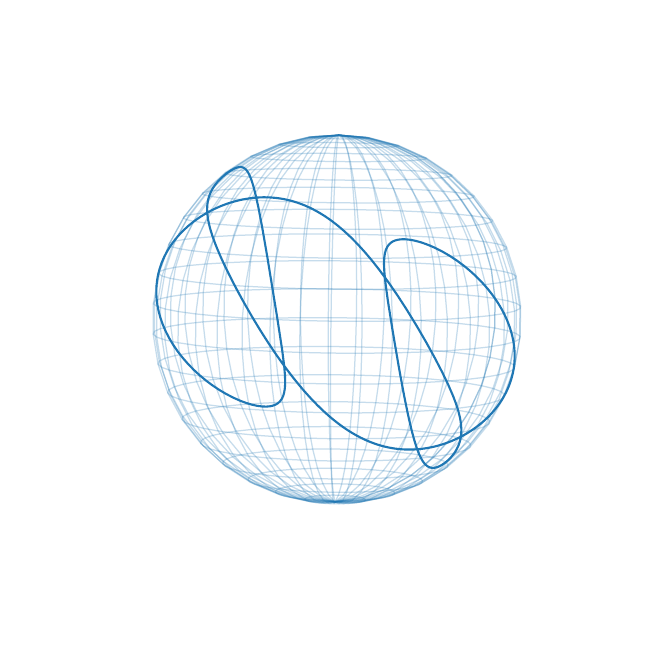}
    \caption{Stable three-cycle in the resonance window at $s=9.3, \omega=1.48$. The quasiperiodic invariant torus is replaced by a periodic attractor consisting of three distinct points in the stroboscopic map. The continuous trajectory closes after three forcing periods.}
    \label{fig:resonance}
\end{figure}

As the forcing amplitude is increased further, individual recurrence
minima continue to deepen until some eventually reach zero. The
corresponding approximate recurrences are then replaced by exact
periodic locking. Figure~\ref{fig:resonance} shows an example of such a
resonant state, in which the stroboscopic invariant curve has collapsed
to three periodic points.

\begin{figure}
    \centering
    \begin{tikzpicture}
      \node[inner sep=0] (resonance) {\includegraphics[width=\columnwidth]{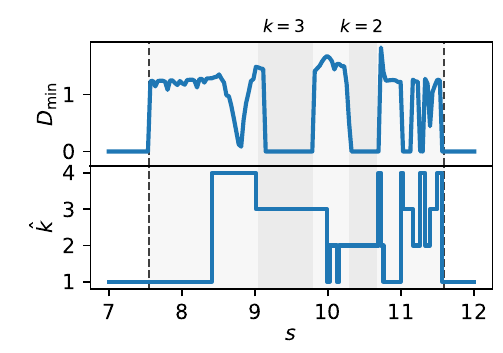}};
      \node[anchor=north west] at ([xshift=4pt,yshift=-4pt]resonance.north west) {\textbf{(a)}};
      \node[anchor=west] at ([xshift=4pt,yshift=-4pt]resonance.west) {\textbf{(b)}};
    \end{tikzpicture}
    \caption{Development of resonance windows with increasing forcing amplitude at fixed driving frequency $\omega=1.48$. (a) Minimum recurrence distance $D_{\min}$. (b) Corresponding recurrence order $\hat{k}=\arg\min_k D_k$. Immediately above the synchronization threshold, the motion is quasiperiodic and all recurrence distances remain finite. With increasing forcing amplitude, isolated resonance windows appear in which $D_{\min}$ reaches numerical zero, indicating exact periodic locking. The shaded regions highlight the dominant low-order resonances $k=3$ and $k=2$. At still larger forcing amplitudes, these resonance windows disappear and the system returns to the synchronized $k=1$ state. Dashed vertical lines indicate the onset of torus dynamics and the return to the synchronized regime.}
    \label{fig:s_scan_resonance}
\end{figure}

To follow the development of these resonances, we vary the forcing
amplitude at fixed driving frequency. The resulting diagnostics are
summarized in figure~\ref{fig:s_scan_resonance}.

The upper panel displays the minimum recurrence distance
\[
D_{\min}
=
\min_{k\le k_{\max}} D_k,
\]
restricted to the lowest recurrence orders. It therefore measures the closest recurrence exhibited by the attractor.
This quantity captures the
dominant resonance hierarchy, since higher-order resonances are typically much
weaker and occupy considerably smaller parameter intervals.
Immediately beyond the synchronization threshold,
$D_{\min}$ remains comparatively large, consistent with the nearly
featureless recurrence spectrum of
figure~\ref{fig:recurrenceDistance_traj}(a). As the forcing amplitude is
increased, $D_{\min}$ decreases and eventually reaches numerical zero
within isolated parameter intervals. These intervals correspond to
exact resonant cycles embedded in the surrounding quasiperiodic regime.

Figure~\ref{fig:s_scan_resonance} reveals a characteristic progression
from synchronized motion to quasiperiodic dynamics, followed by
isolated resonance windows and, at still larger forcing amplitudes, a
return to synchronized motion. We now extend this analysis to the full
two-parameter plane $(\omega,s)$ in order to determine how these resonance windows are organized across parameter space.

\subsection{Global resonance structure and propulsion observables}

\begin{figure}
    \centering
    \panelgraphic[width=\columnwidth]{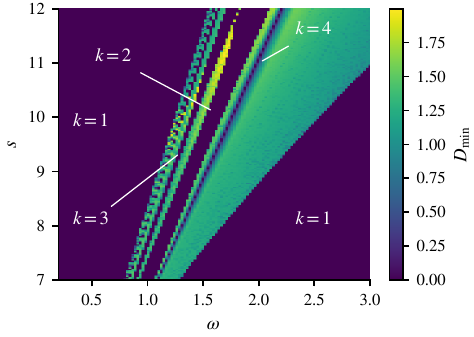}{a}
    \par
    \panelgraphic[width=\columnwidth]{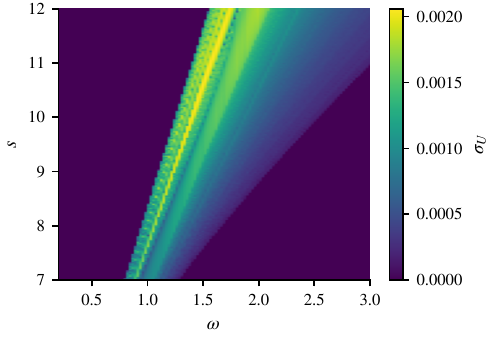}{b}
    \par
    \panelgraphic[width=\columnwidth]{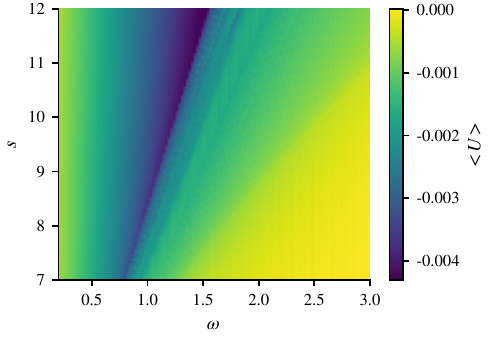}{c}
      \caption{Global organization of the three-mode dynamics in the $(\omega,s)$ parameter plane. (a) Minimum recurrence distance $D_{\min}$, computed from the lowest recurrence orders. Dark regions indicate exact periodic locking, while finite values correspond to quasiperiodic dynamics. Labels denote the primitive recurrence order of the dominant resonance tongues. (b) Standard deviation $\sigma_U$ of the cycle-averaged propulsion velocity. (c) Mean propulsion velocity $\langle U\rangle$.}
    \label{fig:U_2dscan}
\end{figure}

We restrict the global analysis to the lowest recurrence orders, which
already capture the dominant resonance structure.
Figure~\ref{fig:U_2dscan} summarizes the resulting phase diagram
together with two propulsion observables obtained from the same
simulations.

The upper panel displays the minimum recurrence distance
$D_{\min}$, providing a global overview of the synchronization structure. Dark regions correspond to exact periodic locking, whereas
lighter regions indicate quasiperiodic motion. The resonance tongues,
labelled by their primitive recurrence order $k$, identify stable
cycles embedded in the surrounding quasiperiodic regime.

The middle and lower panels show how the underlying shape dynamics is reflected in measurable propulsion observables. From the sequence of cycle-resolved propulsion velocities $\{\bar U_n\}$ we compute the mean propulsion velocity
$
\langle U\rangle=\langle \bar U_n\rangle
$
and the corresponding standard deviation
\[
\sigma_U
=
\sqrt{\left\langle
\left(\bar U_n-\langle U\rangle\right)^2
\right\rangle},
\]
where the brackets denote the average over the cycle index $n$.

The close correspondence between the recurrence diagram and the propulsion
fluctuations demonstrates that the complexity of the underlying shape
dynamics is directly reflected in measurable transport observables.
In particular, the standard deviation $\sigma_U$ of the cycle-resolved
propulsion provides a sensitive experimental signature of synchronization
transitions, quasiperiodicity, and resonant locking, even in parameter
regions where the mean propulsion changes only smoothly.

% The standard deviation is nevertheless only one statistic of the
% cycle-resolved sequence. To expose the temporal organization underlying
% these structures, we next examine the complete signal
% $\{\bar U_n\}$ near a representative transition.

\subsection{Cycle-resolved propulsion as a diagnostic of dynamical transitions}

The global scans show where the fluctuations change but not how they are
organized in time. The complete sequence $\{\bar U_n\}$ resolves this
temporal structure and distinguishes stationary cycle-to-cycle
transport from intermittent propulsion bursts.

\begin{figure}
    \centering
    \panelgraphic[width=0.45\linewidth]{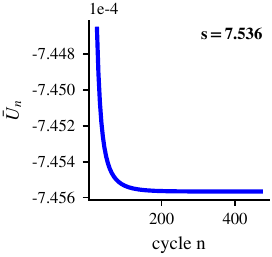}{a}\hfill
    \panelgraphic[width=0.45\linewidth]{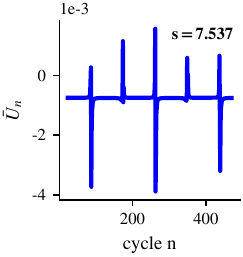}{b}
    \caption{Cycle-resolved propulsion velocity $\bar U_n$ (a) immediately below the onset of intermittency, at $s=7.536$, and (b) immediately above it, at $s=7.537$. Below the transition, $\bar U_n$ converges to a constant value corresponding to period-one synchronization with the driving. Above the transition, rare
propulsion bursts interrupt the nearly periodic sequence. These bursts have
only a minor effect on the mean propulsion but strongly increase the
cycle-to-cycle fluctuations.}
    \label{fig:Un_statistics}
\end{figure}

Figure~\ref{fig:Un_statistics} illustrates the evolution of the
cycle-resolved propulsion signal across the onset of intermittency.
Immediately below the transition, the deformation dynamics is periodic
with the same period as the external driving. After a short transient,
the cycle-resolved propulsion therefore converges to a single constant
value, $\bar U_n=\langle U\rangle$.

Increasing the forcing amplitude from $s=7.536$ to $s=7.537$ produces a
qualitatively different behavior. The trajectory remains close to the
period-one state for long laminar intervals, which are interrupted by
brief intermittent excursions. Although these excursions have only a
small effect on the long-time mean propulsion, they produce pronounced
cycle-to-cycle fluctuations and thereby account for the sharp increase
of $\sigma_U$ observed in the parameter scans of
figure~\ref{fig:U_2dscan}.

This observation also highlights an important difference between
two- and three-mode intermittency. In the two-mode system, the weakly
driven librating state exhibits exactly zero net propulsion, so that
intermittency is readily identified by the appearance of propulsion
bursts and the onset of a nonzero mean velocity. In the three-mode
system, by contrast, the weakly driven state already possesses a finite
background propulsion. Consequently, the accumulated displacement
continues to increase smoothly across the transition, and the onset of
intermittency leaves only a weak signature in the long-time mean
propulsion. Instead, the transition is revealed by the statistics of
the cycle-resolved propulsion signal through the appearance of rare,
large deviations from the otherwise nearly stationary sequence.

The cycle-resolved propulsion signal therefore contains substantially
more information than either its mean value or its standard deviation
alone. Since it can be obtained directly from the displacement over
successive forcing periods, it provides an experimentally accessible
probe of transitions in the underlying deformation dynamics. A
systematic statistical analysis of these time series, including
higher-order moments and temporal correlations, lies beyond the scope
of the present work and is left for future work.

%%%%%%%%%%%%%%%%%%%%%%%%%%%%%%%%%%%%%%%%%%%%%%%%%%%%%%

%%%%%%%%%%%%%%%%%%%%%%%%%%%%%%%%%%%%%%%%%%%%%%%%
\section{Discussion}

The present study shows how membrane constraints can generate nonlinear
shape dynamics and complex propulsion in weakly deformed active
vesicles under periodic forcing. Within the reduced theory, the global
area constraint inherited from local membrane inextensibility is the
source of nonlinearity. It couples the deformation modes and confines
their evolution to compact manifolds in shape space. This mechanism is
sufficient to produce synchronization, quasiperiodicity, resonances,
and intermittency even though the Stokes hydrodynamics, linearized
membrane mechanics, and external forcing are individually linear.
Consequently, harmonic forcing does not need to produce harmonic membrane
motion, and neither the shape dynamics nor the propulsion need to share
the forcing period.

The dynamically selected trajectories in shape space determine the
vesicle displacement through a geometric relation that is quadratic in
the deformation amplitudes. Unlike kinematic swimming theories, the
present framework does not prescribe a deformation
cycle~\cite{ShapereWilczek1987,ShapereWilczek1989}. Instead, the trajectory
is selected by the forced, constrained dynamics, and the resulting
propulsion is a geometric observable of that trajectory. The projected,
torque-based, and phase-space formulations developed here provide
complementary descriptions of this connection between dynamics and
transport.

The two- and three-mode reductions show how the dimension of shape
space changes both the attractors and their transport signatures. The
two-mode system reduces to a periodically driven phase equation and
provides a minimal realization of propulsion onset through a
saddle-node bifurcation on an invariant circle~\cite{Strogatz2014}. Its
mean propulsion is directly proportional to the rotation number.
Retaining a third deformation mode changes this picture qualitatively:
the dynamics evolves on a sphere and supports dynamically selected
invariant tori and resonant cycles.
Moreover, the weakly forced synchronized state already has a finite
mean propulsion. The transition to quasiperiodic torus dynamics is
therefore only weakly expressed in the mean velocity. The mean alone no
longer provides a sufficient characterization of the underlying shape
dynamics.

More generally, our results identify the statistics of the
cycle-resolved propulsion velocity as experimentally accessible
diagnostics of the dynamical state. Its standard deviation already
reveals synchronization transitions and resonance structures that are
only weakly visible in the long-time mean. The complete time series
contains additional information: intermittency appears as long laminar
intervals interrupted by rare propulsion bursts. Distributions of
laminar durations or waiting times between bursts, as well as burst
amplitudes and temporal correlations, may therefore provide a more
detailed characterization of intermittent and quasiperiodic dynamics.
If the membrane shape is resolved simultaneously, trajectories in shape
space and their geometric relation to displacement can also be tested
directly. A systematic analysis of these higher-order statistics and
their robustness to thermal noise, experimental uncertainty, and finite
observation times remains for future work.

The present analysis starts from prescribed time-dependent material
parameters and cortical tractions and focuses on the resulting shape
dynamics and propulsion. Extending the description to the processes
that generate these active forces suggests two directions for future
work. A first step would be to supplement the shape equations with an
autonomous biochemical or cortical oscillator whose dynamics is
independent of the evolving membrane shape. This would generate the
temporal forcing internally while retaining the one-way coupling
assumed here. Encapsulated Min-protein oscillations provide an
experimental example: their reaction--diffusion dynamics produces a
time-dependent redistribution of membrane-bound proteins that drives
autonomous vesicle-shape oscillations
~\cite{Litschel2018,Christ2021}.

A more substantial extension would allow the force-generating process
to depend on the membrane shape. The active fields and the deformation
would then have to be evolved self-consistently, giving rise to true
mechanochemical feedback. Theoretical studies of active
cortex--membrane coupling and curvature-dependent protein recruitment
indicate that such feedback can generate self-sustained oscillations
and propagating waves~\cite{Maitra2014,Wu2018CorticalWaves}. Determining
how these autonomous spatiotemporal dynamics are translated into
propulsion would be a natural continuation of the present work.

Further limitations arise from the restriction to axisymmetric, weakly
deformed vesicles and low-mode truncations. Larger mode sets and
non-axisymmetric deformations may support higher-dimensional invariant
sets, secondary bifurcations, and more complex transport. Stochastic
forcing provides a natural next step for assessing which signatures of
the deterministic attractors persist under experimentally realistic
conditions. More generally, the framework developed here provides a
route towards connecting microscopic active processes, nonlinear shape
dynamics, and macroscopic propulsion within a unified dynamical-systems
description.

%%%%%%%%%%%%%%%%%%%%%%%%%%%%%%%%%%%%%%%%%%%%%%%%%
\appendix 
%%%%%%%%%%%%%%%%%%%%%%%%%%%%%%%%%%%%%%%%%%%%%%%%%

%%%%%%%%%%%%%%%%%%%%%%%%%%%%%%%%%%%%%%%%%%%%%%%%%%%%%%%%%%
\section{Hydrodynamic tractions on a spherical vesicle}
\label{app:lamb_tractions}
%%%%%%%%%%%%%%%%%%%%%%%%%%%%%%%%%%%%%%%%%%%%%%%%%%%%%%%%%%

We work to linear order on the reference sphere $r=a$. The shape is
$r_s=a(1+f)$ with $f(\theta,t)=\sum_{\ell\ge2}f_\ell(t)P_\ell(\mu)$, $\mu=\cos\theta$.
On $r=a$ the kinematic condition gives
\begin{equation}
v_r^\pm(a,\theta)=a\,\dot f(\theta,t)=\sum_{\ell\ge2}a\,\dot f_\ell\,P_\ell(\mu).
\label{eq:kin_vr}
\end{equation}
% Bulk incompressibility together with surface inextensibility on the sphere implies
% \begin{equation}
% \partial_r v_r^\pm(a,\theta)=0.
% \label{eq:drvr0}
% \end{equation}

\subsection{Inextensibility as a normal-derivative condition}
For a surface moving with velocity $\bm v$, local area conservation implies
\begin{equation}
\nabla_s\!\cdot\!\bm v_t + H v_n = 0 ,
\label{eq:surf_inext}
\end{equation}
where $v_n=\bm v\!\cdot\!\bm n$, $\bm v_t=\bm v-v_n\bm n$, and $H$ is the mean curvature.
For an incompressible fluid, $\nabla\!\cdot\!\bm v=0$. Decomposing the divergence near the surface gives
\begin{equation}
\nabla\!\cdot\!\bm v
=
\partial_n v_n + H v_n + \nabla_s\!\cdot\!\bm v_t .
\label{eq:div_decomp}
\end{equation}
Using \eqref{eq:surf_inext} in \eqref{eq:div_decomp} yields
\begin{equation}
\partial_n v_n = 0
\qquad\text{on the membrane}.
\label{eq:drvn}
\end{equation}
Thus, for an incompressible fluid, local membrane inextensibility is equivalent
to the condition that the normal derivative of the normal velocity vanishes at the interface.
On the reference sphere this reduces to 
\begin{equation}
\partial_r v_r(a,\theta)=0
\label{eq:drvr0}
\end{equation}.

\subsection{Lamb solutions}
For each $\ell\ge1$ the axisymmetric (no-swirl) Stokes solutions are written as
\begin{align}
v_r^+(r,\theta)&=(\ell+1)\Bigl(-\frac{a_\ell}{r^{\ell+2}}+\frac{b_\ell}{r^\ell}\Bigr)P_\ell(\mu),
\\
v_\theta^+(r,\theta)&=\Bigl(\frac{a_\ell}{r^{\ell+2}}-\frac{(\ell-2)b_\ell}{r^\ell}\Bigr)\partial_\theta P_\ell(\mu),
\end{align}
in the surrounding fluid  ($r>a$) and
\begin{align}
v_r^-(r,\theta)&=\ell\Bigl(c_\ell r^{\ell-1}+d_\ell r^{\ell+1}\Bigr)P_\ell(\mu),
\\
v_\theta^-(r,\theta)&=\Bigl(c_\ell r^{\ell-1}+\frac{\ell+3}{\ell+1}d_\ell r^{\ell+1}\Bigr)\partial_\theta P_\ell(\mu).
\end{align}
in the interior of the vesicle ($r<a$). The flow field and the stresses for a spherical vesicle can be calculated in three steps. 

Step (i): calculate the coefficients from \eqref{eq:kin_vr} and \eqref{eq:drvr0}.\\
Imposing $v_r^\pm(a)=a\dot f_\ell P_\ell$ and $\partial_r v_r^\pm(a)=0$ yields
\begin{equation}
a_\ell=\frac{\ell}{2(\ell+1)}\,a^{\ell+3}\,\dot f_\ell,
\qquad
b_\ell=\frac{\ell+2}{2(\ell+1)}\,a^{\ell+1}\,\dot f_\ell,
\label{eq:ab_from_fdot}
\end{equation}
\begin{equation}
c_\ell=\frac{\ell+1}{2\ell}\,a^{-(\ell-2)}\,\dot f_\ell,
\qquad
d_\ell=-\frac{\ell-1}{2\ell}\,a^{-\ell}\,\dot f_\ell.
\label{eq:cd_from_fdot}
\end{equation}

Step (ii): From $\nabla p=\eta\nabla^2\bm v$  calculate the pressure.
\begin{equation}
\begin{aligned}
p^+(r,\theta) &=2\eta^+(2\ell-1)\frac{b_\ell}{r^{\ell+1}}P_\ell(\mu),\\
p^-(r,\theta)&=2\eta^-(2\ell+3)\,d_\ell\,r^\ell P_\ell(\mu).
\end{aligned}
\label{eq:p_modes}
\end{equation}

Step (iii): From the flow field and pressure calculate  stress and traction jump.\\
The relevant stress components are
\begin{equation}
\begin{aligned}
\sigma_{rr} & =-p+2\eta\,\partial_r v_r,\\
\sigma_{r\theta} & =\eta\left(\partial_r v_\theta-\frac{v_\theta}{r}+\frac{1}{r}\partial_\theta v_r\right).
\end{aligned}
\label{eq:stress_components}
\end{equation}
Using \eqref{eq:drvr0} gives $\sigma_{rr}^\pm(a)=-p^\pm(a)$.
Substituting \eqref{eq:ab_from_fdot}--\eqref{eq:p_modes} into \eqref{eq:stress_components},
evaluating at $r=a$, and taking the jump
$\llbracket\bm\sigma\cdot\bm n\rrbracket=(\boldsymbol\sigma^+-\boldsymbol\sigma^-)\cdot\bm e_r$
yields the diagonal Legendre form
\begin{align}
e_r\cdot\llbracket\bm\sigma\cdot\bm n\rrbracket
&=-\eta^+\sum_{\ell\ge2} N_\ell\,\dot f_\ell\,P_\ell(\mu),
\\
e_\theta\cdot\llbracket\bm\sigma\cdot\bm n\rrbracket
&=-\eta^+\sum_{\ell\ge2} T_\ell\,\dot f_\ell\,\partial_\theta P_\ell(\mu),
\end{align}
with $\lambda=\eta^-/\eta^+$ and
\begin{equation}
\begin{aligned}
N_\ell &=
\frac{\lambda}{\ell}(\ell-1)(2\ell+3)
+\frac{1}{\ell+1}(\ell+2)(2\ell-1),\\
T_\ell &=\frac{\lambda(\ell-1)+(\ell+2)}{\ell(\ell+1)}.
\end{aligned}
\end{equation}

%=================================================
%\input{NEWHelfrich}
% Appendix B: Linearized Helfrich forces
%================================================

%%%%%%%%%%%%%%%%%%%%%%%%%%%%%%%%%%%%%%%%%%%%%%%%%%%%%%%%%%%%%%
\section{Linearized membrane forces}
\label{app:helfrich_linear}
%%%%%%%%%%%%%%%%%%%%%%%%%%%%%%%%%%%%%%%%%%%%%%%%%%%%%%%%%%%%%%

This appendix derives the linearized membrane tractions entering the
shape evolution equations in the main text. We consider the Helfrich
free energy
\begin{equation}
\mathcal F
=
\int_S
\left[
\frac{\kappa}{2}(H-C)^2
+\kappa_GK
+\gamma
\right]dA ,
\label{eq:app_Helfrich}
\end{equation}
where  $K$ denote  the Gaussian
curvature. As in the main text, the  parameters $M\in \{\kappa, C, \gamma,\, \kappa_G\}$ are decomposed into
homogeneous and small inhomogeneous parts, $M=M_0+\delta M(\bm X)$
and we define the curvature mismatch
\begin{equation}
m=\frac{2}{a}-C_0 .
\end{equation}
Products of the inhomogeneous parts are neglected throughout.

The membrane traction is obtained from the virtual-work principle.
For a virtual displacement
\begin{equation}
\delta\bm X
=
u\,\bm n+\bm\xi ,
\end{equation}
the variation of the free energy defines the force acting on the
membrane,
\begin{equation}
\delta\mathcal F
=
-\int_S
\bm f^{\,\mathrm{on}}
\cdot
\delta\bm X\,dA .
\end{equation}
Throughout this appendix we report the opposite quantity,
$
\bm F^{\rm mem}
=
-\bm f^{\,\mathrm{on}},
$
namely the traction exerted by the membrane on the surrounding fluids, 
which appears in the main text.

%%%%%%%%%%%%%%%%%%%%%%%%%%%%%%%%%%%%%%%%%%%%%%%%%%%%%%%%%%%%%%
\subsection{Linearized geometry}
%%%%%%%%%%%%%%%%%%%%%%%%%%%%%%%%%%%%%%%%%%%%%%%%%%%%%%%%%%%%%%
\label{subsec:geometry}

 The required expansions of geometric quantities $H, K$ and $dA$ for small $f$ about the
reference sphere, which are needed for the calculations are $H=2/a + H_1 + H_2 + O(f^3)$, 
$K=1/a^2+ K_1 + O(f^2)$ and $dA=dA_0 + dA_1 + dA_2$ with
\begin{equation}
\begin{aligned}
H_1 &= -\frac1a(\Delta_0+2)f\\
H_2 & = +\frac2a(f^2+f\Delta_0f)
\end{aligned}
\label{eq:expand_curvature}
\end{equation}
\begin{equation}
\begin{aligned}
dA_1 &= 2f \, dA_0\\
dA_2 & = (f^2+\frac12|\nabla_0f|^2)\, dA_0
\end{aligned}
\label{eq:expand_area}
\end{equation}
and 
\begin{equation}
    K_1=-\frac1{a^2}(\Delta_0+2)f
    \label{eq:expand_Gaussian}
\end{equation}
where $\nabla_0$ and $\Delta_0$ denote the surface gradient and
Laplace--Beltrami operator on the unit sphere. They can be found in references \cite{Vlahovska2026} and \cite{Seifert1999}

The reference radius is chosen such that the undeformed sphere has the
same enclosed volume as the vesicle, implying
\begin{equation}
\int_{S_0}f\,dA_0
=
-\int_{S_0}f^2\,dA_0+O(f^3).
\label{eq:volume_constraint}
\end{equation}
The corresponding first variations,
$\delta_udA=\frac{2u}{a}dA_0$,
$\delta_uH=-a^{-2}(\Delta_0+2)u$, and
$\delta_uK=-a^{-3}(\Delta_0+2)u$,
coincide with the linear terms of
equations~\eqref{eq:expand_area}, \eqref{eq:expand_curvature}, and~\eqref{eq:expand_Gaussian}, since a physical deformation satisfies
$u=af$ to first order. Their role is nevertheless different: the
expansion in $f$ describes the geometry of the deformed membrane,
whereas the virtual displacement $u$ is used to determine the force.

%%%%%%%%%%%%%%%%%%%%%%%%%%%%%%%%%%%%%%%%%%%%%%%%%%%%%%%%%%%%%%
\subsection{Homogeneous membrane}
%%%%%%%%%%%%%%%%%%%%%%%%%%%%%%%%%%%%%%%%%%%%%%%%%%%%%%%%%%%%%%

For homogeneous material parameters,
\begin{equation}
\begin{aligned}
\mathcal F_{\rm hom}
&= \mathcal F_b + \mathcal F_\gamma\\
&=\frac{\kappa_0}{2}\int_S(H-C_0)^2\,dA
+\gamma_0\int_SdA,
\end{aligned}
\end{equation}
where the Gaussian curvature contribution has been omitted because it
is constant for a closed surface.

We expand $(H-C_0)^2dA$ through second order in $f$ using equations~\eqref{eq:expand_curvature} and~\eqref{eq:expand_area}.  
In order to obtain a quadratic form in $f$, all terms linear in $f$ are turned into quadratic terms by using equation~\eqref{eq:volume_constraint}.
Thereby, all terms  proportional to $m$ are canceled. After a partial integration the quadratic bending
energy becomes

\begin{equation*}
    \mathcal F_b^{(2)}
=
\frac{\kappa_0}{2a^2}
\int_{S_0}
f\,\mathcal L_mf\,dA_0,
\end{equation*}
with
\begin{equation}
    \mathcal L_m
=
(\Delta_0+2)
\left(
\Delta_0+2-\frac{(am)^2}{2}
\right).
\label{eq:Lm_operator}
\end{equation}

The tension contribution is obtained by inserting the area element from equation~\eqref{eq:expand_area} and integrating by parts. This gives
\[
\mathcal F_\gamma^{(2)}
=
-\frac{\gamma_0}{2}
\int_{S_0}
f(\Delta_0+2)f\,dA_0
\]

Variation with respect to the normal displacement gives the membrane
traction
\begin{equation}
F_n^{\rm mem}
=
\frac{\kappa_0}{a^3}\mathcal L_mf
-
\frac{\gamma_0}{a}(\Delta_0+2)f.
\label{eq:app_homogeneous_force}
\end{equation}
For homogeneous material parameters there is no tangential membrane
traction.

%%%%%%%%%%%%%%%%%%%%%%%%%%%%%%%%%%%%%%%%%%%%%%%%%%%%%%%%%%%%%%
\subsection{Material inhomogeneities}
%%%%%%%%%%%%%%%%%%%%%%%%%%%%%%%%%%%%%%%%%%%%%%%%%%%%%%%%%%%%%%

We now consider the contribution linear in the spatial variations of
the membrane material parameters,
\begin{equation}
\begin{aligned}
\mathcal F_{\rm inh}
&=
\int_S
\Big[
\frac{\delta\kappa}{2}(H-C_0)^2
-\kappa_0(H-C_0)\delta C\\
&+\delta\kappa_GK
+\delta\gamma
\Big]dA .
\end{aligned}
\label{eq:app_Finh}
\end{equation}
Since equation~\eqref{eq:app_Finh} is already first order in the
inhomogeneous fields, only the linear geometric variations given in
Section~\ref{subsec:geometry} are required. Applying these variations and integrating by
parts yields the normal membrane traction
\begin{align}
F_n^{\rm mem}
={}&
-\frac{m}{a^2}
(\Delta_0+2)\delta\kappa
+\frac{m^2}{a}\delta\kappa
\nonumber\\
&
+\frac{\kappa_0}{a^2}
(\Delta_0+2)\delta C
-\frac{2\kappa_0m}{a}\delta C
\nonumber\\
&
-\frac1{a^3}\Delta_0\delta\kappa_G
+\frac2a\delta\gamma .
\label{eq:app_normal_inh}
\end{align}

Tangential variations leave the membrane shape unchanged and merely
advect the material fields according to
\[
\delta_\xi X
=
-\bm\xi\cdot\nabla_sX,
\qquad
X\in
\{\kappa,\kappa_G,C,\gamma\}.
\]
The corresponding tangential traction is therefore
\begin{equation}
    \begin{aligned}
\bm F_t^{\rm mem}
&=
-\frac{m^2}{2}\nabla_s\delta\kappa
+\kappa_0m\nabla_s\delta C\\
&-\frac1{a^2}\nabla_s\delta\kappa_G
-\nabla_s\delta\gamma .
\end{aligned}
\label{eq:app_tangential_inh}
\end{equation}

%=========================================================
\subsection{Mode representation of tractions}
%==========================================================

We now use the axisymmetric mode notation introduced in the main text.
With
\begin{equation}
g_\ell
=
(\ell-1)(\ell+2),
\qquad
(\Delta_0+2)P_\ell=-g_\ell P_\ell,
\label{eq:app_gell}
\end{equation}
the operator in equation~\eqref{eq:Lm_operator} satisfies
\begin{equation}
\mathcal L_mP_\ell
=
g_\ell
\left[
g_\ell+\frac{(am)^2}{2}
\right]P_\ell.
\end{equation}

We scale lengths by $a$, traction densities by $\kappa_0/a^3$, and
introduce the dimensionless quantities
\begin{equation}
\begin{aligned}
&m\rightarrow am,
\quad
\delta\kappa\rightarrow\frac{\delta\kappa}{\kappa_0},
\quad
\delta C\rightarrow a\delta C,\\
&
\delta\kappa_G\rightarrow
\frac{\delta\kappa_G}{\kappa_0},
\quad
\gamma\rightarrow\frac{a^2\gamma}{\kappa_0}.
\end{aligned}
\end{equation}
The same symbols are used below for the dimensionless variables.
In particular, the dimensionless curvature mismatch becomes $m=2-aC_0$.

The homogeneous radial traction is
\begin{equation}
F_{r,\ell}^{\rm mem,hom}
=
g_\ell
\left(
\beta_\ell+\gamma_0
\right)f_\ell,
\label{eq:app_homogeneous_mode_force}
\end{equation}
where
\begin{equation}
\beta_\ell
=
g_\ell+\frac{m^2}{2}.
\label{eq:app_beta}
\end{equation}

The tractions generated by the inhomogeneous material parameters are
\begin{align}
F_{r,\ell}^{\rm mem,inh}
={}&
(mg_\ell+m^2)\delta\kappa_\ell
-
(g_\ell+2m)\delta C_\ell
\nonumber\\
&+
\ell(\ell+1)\delta\kappa_{G,\ell}
+
2\delta\gamma_\ell,
\label{eq:app_radial_inhomogeneous_mode}
\\
F_{\theta,\ell}^{\rm mem,inh}
={}&
-\frac{m^2}{2}\delta\kappa_\ell
+
m\delta C_\ell
-
\delta\kappa_{G,\ell}
-
\delta\gamma_\ell.
\label{eq:app_tangential_inhomogeneous_mode}
\end{align}
The combination entering the shape equation  is $F^{act, mem}_\ell=F_{r,\ell}^{\rm mem,inh}
+
2F_{\theta,\ell}^{\rm mem,inh}$, which becomes
\begin{equation}
\begin{aligned}
F^{act, mem}_\ell
=
g_\ell
\Big(
m\delta\kappa_\ell
-
\delta C_\ell
+
\delta\kappa_{G,\ell}
\Big)
\end{aligned}
\label{eq:app_combined_inhomogeneous_mode}
\end{equation}
and is used in this form in the main text.
The cancellation of the inhomogeneous tension is a consequence of
local inextensibility. Moreover, the prefactor $g_\ell$ makes
the membrane-bound contribution vanish for $\ell=1$, as required by
translational invariance.

%==========================================================================

%%%%%%%%%%%%%%%%%%%%%%%%%%%%%%%%%%%%%%%%%%%%%%%%%%%%%%%%%%
\section{Lyapunov property of the quadratic elastic energy}
\label{app:Lyapunov}
%%%%%%%%%%%%%%%%%%%%%%%%%%%%%%%%%%%%%%%%%%%%%%%%%%%%%%%%%%%

For vanishing drive ($s=0$), the projected dynamics reads
\begin{equation}
\dot{\bm q}
=
-
\bm P_\Gamma(\bm q)\,
\hat{\bm\Gamma}\hat{\bm\beta}\,\bm q,
\qquad
\bm P_\Gamma
=
\bm I
-
\frac{(\hat{\bm\Gamma}\bm q)\otimes\bm q}
{\bm q\cdot(\hat{\bm\Gamma}\bm q)}.
\label{eq:undriven_appendix}
\end{equation}

Consider the quadratic elastic energy
\begin{equation}
\mathcal E(\bm q)
=
\frac12\,\bm q\cdot\hat{\bm\beta}\bm q,
\label{eq:elastic_energy_appendix}
\end{equation}
whose gradient is
$\nabla_{\bm q}\mathcal E=\hat{\bm\beta}\bm q$.
Denoting
\[
\bm x:=\hat{\bm\beta}\bm q,
\qquad
\bm Q:=\hat{\bm\Gamma}\bm q,
\qquad
\mu:=\bm q\cdot\bm Q,
\]
the energy dissipation along trajectories becomes
\begin{align}
\frac{d\mathcal E}{dt}
&=
\nabla\mathcal E\cdot\dot{\bm q}
\nonumber\\
&=
-
\bm x\cdot\hat{\bm\Gamma}\bm x
+
\frac{(\bm x\cdot\bm Q)^2}{\mu}.
\label{eq:dE_intermediate}
\end{align}

Introduce the $\hat{\bm\Gamma}$–weighted inner product
\[
\langle \bm a,\bm b\rangle_\Gamma
:=
\bm a\cdot\hat{\bm\Gamma}\bm b .
\]
Then
\[
\bm x\cdot\hat{\bm\Gamma}\bm x
=
\langle\bm x,\bm x\rangle_\Gamma,
\qquad
\bm x\cdot\bm Q
=
\langle\bm x,\bm q\rangle_\Gamma,
\qquad
\mu
=
\langle\bm q,\bm q\rangle_\Gamma .
\]
Equation~\eqref{eq:dE_intermediate} therefore reads
\[
\frac{d\mathcal E}{dt}
=
-
\langle\bm x,\bm x\rangle_\Gamma
+
\frac{\langle\bm x,\bm q\rangle_\Gamma^2}
{\langle\bm q,\bm q\rangle_\Gamma}.
\]
By the Cauchy–Schwarz inequality in the
$\Gamma$–weighted inner product,
\[
\langle\bm x,\bm x\rangle_\Gamma
\,
\langle\bm q,\bm q\rangle_\Gamma
\ge
\langle\bm x,\bm q\rangle_\Gamma^2,
\]
and hence
\begin{equation}
\frac{d\mathcal E}{dt}
\le 0.
\end{equation}

Equality holds only when $\bm x$ and $\bm q$ are
collinear in the $\Gamma$–metric,
which corresponds to stationary points of the elastic torque.
Thus the quadratic Helfrich energy
acts as a Lyapunov function for the undriven dynamics.

%%%%%%%%%%%%%%%%%%%%%%%%%%%%%%%%%%%%%%%%%%%%%%%%%%%%%%%%%%%%%%

%%%%%%%%%%%%%%%%%%%%%%%%%%%%%%%%%%%%%%%%%%%%%%%%%%%%%%%%%%%%%
\section{Saddle-node bifurcation}
\label{app:snic_map}
%%%%%%%%%%%%%%%%%%%%%%%%%%%%%%%%%%%%%%%%%%%%%%%%%%%%%%%%%%%%%%

In this Appendix we collect some well-known results for saddle-node bifurcations in dynamical systems on a circle for the reader's convenience. For more details see \cite{KatokHasselblatt1995, guckenheimer_holmes, arnold}.
The reduced phase dynamics is governed by a $T$–periodic ordinary differential equation of the form
\begin{equation}
\dot\psi = f(\psi,t;s),
\quad
f(\psi,t+T;s)=f(\psi,t;s),
\label{eq:scalarODE}
\end{equation}
where $T=2\pi/\omega$ and $s$ denotes the driving amplitude. We define the associated stroboscopic (Poincar\'e) map
\begin{equation}
\mathcal P_s(\psi_0) := \psi(T;\psi_0,s),
\end{equation}
where $\psi(T;\psi_0,s)$ is the solution at time $T$ starting from $\psi_0$ at $t_0=0$.
%with $f$ continuous in $t$ and locally Lipschitz in %$\psi$.  
 If not stated otherwise, the range of $\psi$ is considered here as the real axis.  A fixed point of the stroboscopic map  $\mathcal P(\psi^*)=\psi^*$ corresponds to a $T$–periodic solution
$\psi(t+T, \psi_0)=\psi(t,\psi_0)$, i.e. phase locking to the external forcing.

\subsection{Equivalence of zero rotation number and phase locking}

Because equation~\eqref{eq:scalarODE} is scalar and solutions are unique, trajectories cannot cross.  
Thus, if $\psi_1<\psi_2$ initially, then
$
\psi(t;\psi_1)<\psi(t;\psi_2)$
for all $t>0$.

Consequently, the map $\mathcal{P}$ is strictly increasing:
\[
\psi_1<\psi_2
\;\Rightarrow\;
\mathcal P_s(\psi_1)<\mathcal P_s(\psi_2).
\]

In the main text, we made use of the fact, that a vanishing rotation number, $\rho=0$ is equivalent to the existence of a fixed point $\mathcal P(\psi^*)=\psi^*$. 
Let us consider  both directions of this proposition: 
%%%%%%%%%%%%%%%%%%%%%%%%%%%%%%%%%%%%%%%%%%%%%%%%%%%%%%%%%%%

(i)
Assume $\rho=0$ but $\mathcal P$ has no fixed point.  
Since $\mathcal P$ is continuous and strictly increasing, either
$
\mathcal P_s(\psi)>\psi
\quad\text{for all }\psi,
\quad\text{or}\,
\mathcal  P_s(\psi)<\psi
\quad\text{for all }\psi.
$
In the first case, $\mathcal P_s^{\,n}(\psi_0)-\psi_0 \ge n\varepsilon$ for some $\varepsilon>0$,
implying $\rho>0$; in the second case $\rho<0$.  
Both contradict $\rho=0$.  
Thus a fixed point must exist.

(ii)  
If the stroboscopic map $\mathcal P_s$ possesses a fixed point $\psi^*$,
$\mathcal P_s(\psi^*)=\psi^*$, then $\mathcal P_s^n(\psi^*)=\psi^*$ for all $n$ and the
rotation number computed along this orbit vanishes. Since $\mathcal P_s$ is strictly monotone, the rotation number exists and is independent
of the initial condition~\cite{KatokHasselblatt1995,arnold}. It therefore follows
that $\rho=0$ for all trajectories, corresponding to phase locking and the absence
of net drift.
\hfill$\square$
\subsection{Stroboscopic map and bifurcation conditions}
A propelling–to–locked transition corresponds to a saddle--node bifurcation of the orientation-preserving circle map $\mathcal P_s$. At the critical driving amplitude $s=s^*$ there exists a marginal fixed point $\psi=\psi^*$ satisfying
\begin{equation}
\begin{aligned}
\mathcal P_{s^*}(\psi^*)=\psi^*,
\qquad
\partial_\psi \mathcal P_{s^*}(\psi^*)=1,\\
\qquad
\partial_{\psi\psi}\mathcal P_{s^*}(\psi^*)\neq 0,
\qquad
\partial_s \mathcal P_{s^*}(\psi^*)\neq 0.
\label{eq:snic_conditions}
\end{aligned}
\end{equation}
These conditions define a saddle--node on an invariant circle (SNIC) bifurcation.

\subsection{Local normal form of the stroboscopic map}

We expand the map near the bifurcation point by introducing
\begin{equation}
x_n = \psi_n-\psi^*,
\qquad
\varepsilon = s-s^*.
\end{equation}
A Taylor expansion of $\mathcal P_s$ around $(\psi^*,s^*)$ yields
\begin{equation}
x_{n+1}
=
x_n
+
\alpha\,\varepsilon
+
\beta\,x_n^2
+
\mathcal O(\varepsilon x_n,x_n^3,\varepsilon^2),
\label{eq:map_normal_form}
\end{equation}
where
\begin{equation}
A = \left.\partial_s \mathcal P_s(\psi)\right|_{(\psi^*,s^*)},
\quad
B = \frac12\left.\partial_{\psi\psi}\mathcal P_s(\psi)\right|_{(\psi^*,s^*)}.
\end{equation}
The absence of a linear term in $x_n$ reflects the marginal stability condition $\partial_\psi\mathcal P_{s^*}(\psi^*)=1$.

Equation~\eqref{eq:map_normal_form} is the universal normal form of a saddle--node bifurcation for an orientation-preserving circle map.

\subsection{Continuous approximation and SNIC scaling}

Close to the bifurcation, the change of $x_n$ per iteration is small, and the discrete map may be approximated by a continuous equation in the iteration index $n$,
\begin{equation}
\frac{dx}{dn}
=
A\,\varepsilon
+
B\,x^2.
\label{eq:flow_from_map}
\end{equation}
This equation captures the slow passage through the bottleneck near $x=0$. After rescaling $x$ and $n$, equation~\eqref{eq:flow_from_map} reduces to the canonical SNIC normal form
\begin{equation}
\frac{dx}{dn} = \varepsilon + x^2.
\end{equation}

The time $\tau_n$ spent in the bottleneck (measured in number of forcing periods) diverges as
\begin{equation}
\tau_n \sim \varepsilon^{-1/2}.
\end{equation}
Since each traversal of the bottleneck produces a finite phase advance, the mean phase drift per forcing period (rotation number) scales as
\begin{equation}
\rho \sim \varepsilon^{1/2}.
\end{equation}

%%%%%%%%%%%%%%%%%%%%%%%%%%%%%%%%%%%%%%%%%%%%%%%%%%%%%%%%%%%%%%%%%%%%
\section{Explicit torque components in spherical coordinates}
\label{eq:3modeappendix}
%%%%%%%%%%%%%%%%%%%%%%%%%%%%%%%%%%%%%%%%%%%%%%%%%%%%%%%%%%%%%%%%%%%%%

In this Appendix we evaluate the torque components entering the angular
equations derived in the main text. Using
$
q_2=\cos\psi,$
$q_3=\sin\psi\cos\chi,$
$q_4=\sin\psi\sin\chi,$
the metric vector
$
\bm Q=\hat{\bm \Gamma}\bm q,
$
has components
\begin{equation}
\begin{aligned}
&Q_2=\Gamma_2\cos\psi,\\
&Q_3=\Gamma_3\sin\psi\cos\chi,\\
&Q_4=\Gamma_4\sin\psi\sin\chi.
\end{aligned}
\end{equation}
The scalar factor appearing in equation~\eqref{eq:tau} is therefore
\begin{equation}
\begin{aligned}
\mu
&=
\bm q\cdot\bm Q\\
=&
\Gamma_2\cos^2\psi
+
\Gamma_3\sin^2\psi\cos^2\chi
+
\Gamma_4\sin^2\psi\sin^2\chi.
\end{aligned}
\end{equation}

\subsection{Torque decomposition}

Using the definition of the projected force from the main text,
\begin{equation}
\bm F^p
=
\hat{\bm\Gamma}\hat{\bm\beta}\bm q
+
\hat{\bm\Gamma}\bm F,
\end{equation}
and noting that the diagonal matrices
$\hat{\bm\Gamma}$ and $\hat{\bm\beta}$ commute, we may write
\begin{equation}
\bm F^p
=
\hat{\bm\beta}\bm Q+\hat{\bm\Gamma}\bm F.
\end{equation}
The torque in equation~\eqref{eq:tau} therefore separates as
\begin{equation}
\bm\tau
=
\bm\tau^{\rm int}
+
\bm\tau^{\rm drive},
\end{equation}
where
\begin{equation}
\bm\tau^{\rm int}
=
\bm Q\times(\hat{\bm\beta}\bm Q),
\qquad
\bm\tau^{\rm drive}
=
\bm Q\times(\hat{\bm\Gamma}\bm F).
\end{equation}

\subsection{Intrinsic torque}

The intrinsic torque is
\begin{equation}
\bm\tau^{\rm int}
=
\bm Q\times(\hat{\bm\beta}\bm Q)
=
\begin{pmatrix}
Q_3Q_4(\beta_4-\beta_3)\\
Q_4Q_2(\beta_2-\beta_4)\\
Q_2Q_3(\beta_3-\beta_2),
\end{pmatrix}.
\label{eq:tau-int-cartesian}
\end{equation}

which becomes
\begin{align}
\tau^{\rm int}_2
&=
\Gamma_3\Gamma_4(\beta_4-\beta_3)
\sin^2\psi\,\sin\chi\cos\chi,
\\
\tau^{\rm int}_3
&=
\Gamma_2\Gamma_4(\beta_2-\beta_4)
\sin\psi\cos\psi\,\sin\chi,
\\
\tau^{\rm int}_4
&=
\Gamma_2\Gamma_3(\beta_3-\beta_2)
\sin\psi\cos\psi\,\cos\chi.
\end{align}

The tangent components are obtained using
\begin{equation}
\begin{aligned}
&\tau_\chi=-\tau_3\sin\chi+\tau_4\cos\chi,\\
&\tau_\psi
=
-\tau_2\sin\psi
+\tau_3\cos\psi\cos\chi
+\tau_4\cos\psi\sin\chi.
\end{aligned}
\end{equation}
For the intrinsic torque, this yields
\begin{equation}
\begin{aligned}
\tau_\chi^{\rm int}
={}&
\Gamma_2\sin\psi\cos\psi
\Big[
\Gamma_3(\beta_3-\beta_2)\cos^2\chi
\\
&\qquad
+\Gamma_4(\beta_4-\beta_2)\sin^2\chi
\Big],
\end{aligned}
\label{eq:tau-chi-int}
\end{equation}
and
\begin{equation}
\begin{aligned}
&\tau_\psi^{\rm int}
=\, 
\sin\psi\,\sin\chi\cos\chi
\Big\{
\Gamma_3\Gamma_4(\beta_3-\beta_4)\sin^2\psi
\\
&
+\Gamma_2\cos^2\psi
\left[
\Gamma_4(\beta_2-\beta_4)
+\Gamma_3(\beta_3-\beta_2)
\right]
\Big\}.
\end{aligned}
\label{eq:tau-psi-int}
\end{equation}

\paragraph{Driven torque in spherical coordinates.}

For
\[
\bm F=(F_2,F_3,F_4),
\]
the driven torque
\[
\bm\tau^{\rm drive}
=
\bm Q\times(\hat{\bm\Gamma}\bm F)
\]
has Cartesian components
\begin{align}
\tau^{\rm drive}_2
&=
\Gamma_3\Gamma_4\,\sin\psi
(\cos\chi\,F_4-\sin\chi\,F_3),
\\
\tau^{\rm drive}_3
&=
\Gamma_2\Gamma_4
(\sin\psi\,\sin\chi\,F_2-\cos\psi\,F_4),
\\
\tau^{\rm drive}_4
&=
\Gamma_2\Gamma_3
(\cos\psi\,F_3-\sin\psi\,\cos\chi\,F_2).
\end{align}
Projection onto the tangent basis yields
\begin{equation}
\begin{aligned}
\tau_\chi^{\rm drive}
={}&
\Gamma_2\Gamma_3
\cos\psi\cos\chi\,F_3
+
\Gamma_2\Gamma_4
\cos\psi\sin\chi\,F_4
\\
&\quad
-
\Gamma_2\sin\psi
\Bigl[
\Gamma_3\cos^2\chi
+
\Gamma_4\sin^2\chi
\Bigr]
F_2,
\end{aligned}
\end{equation}
and
\begin{equation}
\begin{aligned}
\tau_\psi^{\rm drive}
={}&
\Gamma_2(\Gamma_4-\Gamma_3)
\sin\psi\cos\psi
\sin\chi\cos\chi\,F_2
\\
&\quad
+
\Gamma_3\sin\chi
\Bigl(
\Gamma_4\sin^2\psi
+
\Gamma_2\cos^2\psi
\Bigr)
F_3
\\
&\quad
-
\Gamma_4\cos\chi
\Bigl(
\Gamma_3\sin^2\psi
+
\Gamma_2\cos^2\psi
\Bigr)
F_4.
\end{aligned}
\end{equation}

% %\include{constants}
\bibliographystyle{iopart-num}
\bibliography{active_vesicle}
\end{document}